\documentclass[11pt,a4paper]{article}
\usepackage[utf8x]{inputenc}
\usepackage[T1]{fontenc}

\usepackage[pdftex]{graphicx} 
\usepackage[pdftex,linkcolor=black,pdfborder={0 0 0}]{hyperref} 
\usepackage{calc} 
\usepackage{enumitem} 
\usepackage{cite}

\usepackage[a4paper, lmargin=0.1666\paperwidth, rmargin=0.1666\paperwidth, tmargin=0.1111\paperheight, bmargin=0.1111\paperheight]{geometry} 

\usepackage[all]{nowidow} 
\usepackage[protrusion=true,expansion=true]{microtype} 

\usepackage{amssymb}
\usepackage{amsmath}

\hypersetup{
pdfsubject = {},
pdftitle = {},
pdfauthor = {}
}

\usepackage{import}

\usepackage{amsfonts}
\usepackage{amscd}
\usepackage{amssymb}
\usepackage{amsmath,bbm}
\usepackage{graphicx}
\usepackage{subcaption}

\usepackage{epsfig}
\usepackage{latexsym}
\usepackage{mathtools}
\usepackage{hyperref}
\usepackage[vcentermath]{youngtab}
\usepackage{mathrsfs}
\usepackage{url}
\usepackage{comment}
\usepackage[svgnames]{xcolor}
\usepackage{tikz}
\usepackage{tikz-cd}
\usetikzlibrary{arrows,arrows.meta,3d,calc}

\usepackage{amsmath}
\usepackage{mathtools}
\numberwithin{equation}{section}
\usepackage{slashed}
\usepackage{braket}

\usepackage{tikz-feynman}
\makeatletter
\renewcommand{\tikzfeynman@luatex@required@path}{}
\renewcommand{\tikzfeynman@luatex@required@key}{}
\makeatother

\usepackage{cancel}
\usepackage{parskip}

\hypersetup{colorlinks=true, linkcolor=blue,citecolor=blue,filecolor=black,urlcolor=black,}

\tikzcdset{arrow style=tikz, diagrams={>=stealth}}

\newcommand\cB{\mathcal{B}}

\newcommand\cI{\mathcal{I}}

\newcommand\cL{\mathcal{L}}
\newcommand\cM{\mathcal{M}}

\newcommand\cO{\mathcal{O}}

\newcommand\cQ{\mathcal{Q}}
\newcommand\cR{\mathcal{R}}

\newcommand\cU{\mathcal{U}}

\newcommand\cW{\mathcal{W}}

\newcommand\bC{\mathbb{C}}

\newcommand\bN{\mathbb{N}}

\newcommand\bQ{\mathbb{Q}}
\newcommand\bR{\mathbb{R}}

\newcommand\bZ{\mathbb{Z}}

\newcommand\Hom{\mathrm{Hom}}

\newcommand{\dbar}{\mathchar'26\mkern-11mu d} 

\newcommand{\brap}[1]{{\left( {#1} \right)}}
\newcommand{\bras}[1]{{\left[ {#1} \right]}}
\newcommand{\brac}[1]{{\left\{ {#1} \right\}}}

\newcommand{\brav}[1]{{\left| {#1} \right|}}

\definecolor{cardinal}{rgb}{0.6,0,0}
\definecolor{darkgreen}{rgb}{0,0.5,0}
\definecolor{golden}{rgb}{0.92, 0.7, 0}
\definecolor{midnight}{rgb}{0, 0, 0.5}
\definecolor{darkblue}{rgb}{0.2, 0, 0.8}

\begin{document}

\begin{titlepage}
\begin{flushright}

\end{flushright}

\vskip 3cm

\begin{center}

{\huge Decompactification Limits of Non-Compact Gauge Theory }

\vskip 1cm
Finn Gagliano and Christopher Tudball
\vskip 0.5cm

\begin{tabular}{ll}
 & Department of Mathematical Sciences, Durham University, UK
\end{tabular}

\vskip 1cm

\end{center}

\noindent
\begin{center}
    \textbf{Abstract:}
\end{center}
The required absence of global symmetries in quantum gravity has been used to imply that all non-compact gauge theories are in the swampland. This argument stems from the idea that non-compact gauge symmetries always seem to be accompanied by global symmetries that cannot be broken with a finite number of fields. In this work, we investigate whether these symmetries can be broken by an \textit{uncountable infinity} of fields. We find that the symmetries can be broken, but as soon as we add these fields the EFT breaks down and in some cases decompactifies to a higher-dimensional theory without the non-compact gauge symmetry, akin to undoing a Kaluza-Klein reduction on a non-compact space. We make various comments on the species scale, free parameters, and the Weak Gravity Conjecture along the way.

\raggedright
\small
\vspace*{\fill}
\noindent\rule{0.4\textwidth}{0.4pt} \\
finn.gagliano@durham.ac.uk \\
christopher.a.tudball@durham.ac.uk

\end{titlepage}

\setcounter{tocdepth}{3}
\tableofcontents

\section{Introduction}\label{sec:introduction}

In \cite{Vafa:2005ui} it was observed that a seemingly consistent effective field theory (EFT) may be rendered inconsistent when we try to extend it to a theory of quantum gravity.
From this idea grew the swampland programme, with the aim of identifying conditions that EFTs must satisfy to be in the `landscape', the space of consistent theories of quantum gravity. If an EFT does not satisfy all of these conditions, then it is in the `swampland' -- it cannot be consistently coupled to quantum gravity. See \cite{Rudelius:2024vmc,vanBeest:2021lhn,Rudelius:2024mhq,Harlow:2022ich,Palti:2019pca,Brennan:2017rbf,Grana:2021zvf,Reece:2023czb} for introductions and reviews of the swampland programme.

One of the central swampland conditions is the idea that theories in the landscape never exhibit global symmetries \cite{Hawking:1975vcx,Zeldovich:1976vw,Zeldovich:1977mk,Banks:1988yz,Giddings:1988cx,Abbott:1989jw,Coleman:1989zu,Kallosh:1995hi,Banks:2010zn,Harlow:2018tng,Harlow:2020bee,Chen:2020ojn,Hsin:2020mfa,Yonekura:2020ino,McNamara:2021cuo}, including generalised global symmetries \cite{Gaiotto:2014kfa}. Any global symmetry of an EFT must eventually be explicitly broken before we extend the EFT to probe the scales at which quantum gravitational effects become relevant, or alternatively the theory must be replaced with one in which the global symmetry is gauged away.
Even when a symmetry is gauged, this condition still has teeth to bare, for in general a gauge symmetry comes with associated (higher-form) global symmetries which must themselves be explicitly broken or gauged.

For example, pure U(1) gauge theory exhibits a pair of global symmetries, called the electric 1-form symmetry and the magnetic $(d-3)$-form symmetry. The charged objects under these symmetries are the Wilson lines and `t Hooft defects, respectively. Upon adding a charge 1 electron and charge 1 magnetic monopole to the theory, these symmetries are explicitly broken. Another swampland condition is that in the landscape, whenever we have a gauge symmetry, we should have operators transforming in every representation of the gauge group -- a so called ``complete spectrum'', where generally these operators are there to break the higher-form symmetries associated to the gauge group \cite{Rudelius:2020orz,Heidenreich:2021xpr}. We will review how this works in §\ref{sec:background}.

A corollary of these two swampland constraints is that gauge groups must then be compact. There are a number of arguments for this, but taking the example of $\bR$ gauge theory, not only do we have the usual 1-form global symmetry that cannot be broken with a finite number of fields, but we also have additional 0-form symmetries wherein any matter fields of relatively irrational charge transform differently \cite{Banks:2010zn,Harlow:2018tng}.

In this work, we demonstrate a loophole in these arguments. Namely, if we add an uncountable infinity of interacting fields, both of these global symmetries are explicitly broken, and the spectrum is complete, seemingly saving the $\bR$ gauge theory from the swampland. However, an EFT with uncountably infinitely many fields is troubling, and the scale at which quantum gravitational effects becomes important, the species scale, vanishes in Planck units so that our EFT is invalid for all energy scales.

However, we show how this theory with uncountably infinitely many fields at certain restricted points of parameter space can be decompactified to a theory of finitely many fields in one spacetime dimension higher, in which the $\bR$ gauge symmetry becomes a part of the diffeomorphisms of this higher dimensional theory, akin to a Kaluza-Klein reduction on $\bR$.

This work is organised as follows.
In §\ref{sec:background} we review the background material to aid understanding of this work and review the existing arguments for why continuous non-compact gauge theories should be in the swampland, and why continuous compact gauge theories avoid these arguments.
The main new results lie in §\ref{sec:breaking-symmetries}, where we explain how to break the global symmetries by adding uncountably infinitely many fields, and in §\ref{sec:extra-dimensions}, where we explain how the $\bR$ gauge theory with uncountably infinitely many fields decompactifies to a theory with finitely many fields in one higher spacetime dimension without a $\bR$ gauge symmetry.
Lastly, we provide concluding remarks and future directions in §\ref{sec:conclusion}.
We also include some appendices -- Appendices \ref{appendix:other-methods-of-integration} and \ref{appendix:finite-charge-interval} detail alternatives to some of the steps taken in the many body of the text, Appendix \ref{appendix:extra-dim-propagators-and-species-scale} gives details of the vanishing species scale, and Appendix \ref{appendix:matching-mass-dimensions} describes the matching of mass dimensions across the decompactification limit.

\section{Background} \label{sec:background}

As discussed, EFTs with a generalised global symmetry are expected to be in the swampland. A theory possesses a $p$-form symmetry if one can define codimension-$(p+1)$ topological operators $\cU(\Sigma_{d-p-1})$, meaning correlation functions of the theory depend only on the topology of $\Sigma$. Such symmetries act on $p$-dimensional operators of the theory. One can explicitly break such a symmetry by allowing the charged $p$-dimensional operators to end on some $(p-1)$-dimensional operator transforming in the same representation of the gauge group. Doing so means that correlation functions depend non-topologically on $\Sigma_{d-p-1}$, as in Figure~\ref{fig:breaking}, such that $\cU(\Sigma_{d-p-1})$ is no longer topological. This means that completing the spectrum is the same as breaking all (potentially non-invertible) generalised global symmetries \cite{Rudelius:2020orz,Heidenreich:2021xpr}.

\begin{figure}
    \centering
    \begin{tikzpicture}[scale=1.2, thick]

    \begin{scope}
        \draw[red] (0,-1.2) -- (0,1.2);
        \node[red, right] at (0,1.1) {$\cW$};

        \draw[blue]
          (0.8,0) arc[start angle=0, end angle=80,  x radius=0.8, y radius=0.4];
        \draw[blue]
          (-0.8,0) arc[start angle=180, end angle=360, x radius=0.8, y radius=0.4];
        \draw[blue]
          (0.8,0) arc[start angle=360, end angle=100, x radius=0.8, y radius=0.4];

        \node[blue, left] at (-0.8,0.5) {$\cU(\Sigma)$};
      \end{scope}

      \draw[->] (1.5,0) -- (2.7,0);

      \begin{scope}[xshift=4cm]
        \draw[red] (0,-1.2) -- (0,1.2);
        \node[red, right] at (0,1.1) {$\widetilde{\cW}$};

        \draw[blue] (1.3,0) ellipse [x radius=0.8, y radius=0.4];
        \node[blue, right] at (2.1,0.5) {$\cU(\Tilde{\Sigma})$};
      \end{scope}


      \begin{scope}[yshift=-2.8cm]

        \begin{scope}
          \draw[red] (0,-0.45) -- (0,0.9);
          \node[red, right] at (0,0.8) {$\cW$};
          \node at (0,-0.45) {$\bullet$};
          \node[right] at (0,-0.45) {$\cO$};
          \begin{scope}[yshift=0.2cm]
            \draw[blue]
              (0.8,0) arc[start angle=0, end angle=80,  x radius=0.8, y radius=0.4];
            \draw[blue]
              (-0.8,0) arc[start angle=180, end angle=360, x radius=0.8, y radius=0.4];
            \draw[blue]
              (0.8,0) arc[start angle=360, end angle=100, x radius=0.8, y radius=0.4];

            \node[blue, left] at (-0.8,0.5) {$\cU(\Sigma)$};
          \end{scope}
        \end{scope}

        \draw[->] (1.5,0) -- (2.7,0);

        \begin{scope}[xshift=4cm]
          \draw[red] (0,-0.45) -- (0,0.9);
          \node[red, right] at (0,0.8) {$\cW$};
          \node at (0,-0.45) {$\bullet$};
          \node[right] at (0,-0.45) {$\cO$};
          \begin{scope}[yshift=-0.0cm]
            \draw[blue] (1.3,0) ellipse [x radius=0.8, y radius=0.4];
            \node[blue, right] at (2.1,0.5) {$\cU(\Tilde{\Sigma})$};
          \end{scope}
        \end{scope}

    \end{scope}
    \end{tikzpicture}
    \caption{\textbf{Top left to top right:} The blue line, labelled $\cU(\Sigma)$, is a topological operator generating a symmetry action on the charged operator, the red line labelled $\cW$, producing a new line $\widetilde{\cW}$. \\ \textbf{Bottom left to bottom right:} End-ability of $\cW$ on the operator $\cO$ means correlation functions depend non-topologically on $\Sigma$, as we can go from $\Sigma$ to $\Tilde{\Sigma}$ smoothly without intersecting $\cW$, breaking the symmetry.}
    \label{fig:breaking}
\end{figure}

As mentioned, a consequence of the no global symmetries conjecture is that gauge groups must be compact \cite{Banks:2010zn} -- non-compact gauge theories always seem to be accompanied by a global symmetry. Consider the action of $G=\bR$ gauge theory
\begin{equation}
    S = \int_{\cM_d} \frac{1}{2g^2} F \wedge * F \,.
\end{equation}
We can define operators called Wilson lines,
\begin{equation}
    \cW_q(\gamma) = e^{iq\int_\gamma A}\,,
\end{equation}
where $q\in \bR$. When $\partial\gamma=\varnothing$, we have that under a small gauge transformation
\begin{equation}
    \int_\gamma A \rightarrow \int_\gamma (A  + d\lambda) = \int_\gamma A
\end{equation}
by Stokes' theorem. Therefore, there are no constraints on $q$ so that all $\cW_q$ are gauge-invariant operators and there are no local operators on which these operators can end. That is, we can understand this as the presence of an $\bR^{[1]}$ symmetry, under which the Wilson lines are charged.

Compare this with the Wilson lines of $G=U(1)$. In this theory we also have large gauge transformations,
\begin{equation}
    \int_\gamma A \rightarrow\int_\gamma (A + 2\pi n) = (\int_\gamma A) + 2\pi n(\gamma) \,,
\end{equation}
where $n$ is a closed but not exact 1-form that integrates to $n(\gamma)\in\bZ$. Then, for Wilson lines $\cW_q$ to be invariant under such a transformation, we require $q\in \bZ$. So, the only gauge invariant Wilson lines of the theory are those with integer charge. Again, there are no local operators on which the lines can end, and so the Wilson lines are charged under a $U(1)^{[1]}$ symmetry.

We can define the associated topological operator in each case to be
\begin{equation}
    \cU_\alpha(\Sigma_{d-2}) = e^{i\alpha \int *F} \,,
\end{equation}
where $d*F=0$ can be seen as the conservation of a 2-form current, with $\alpha \in U(1) $ or $ \bR$, depending on the choice of gauge group. This operator measures the charge of the Wilson lines:
\begin{equation}
    \cU_\alpha(\Sigma) \cW_q (\gamma) = e^{i\alpha q \mathrm{Link(\Sigma, \gamma)}} \cW_q(\gamma)
\end{equation}
as would be the case for an ordinary symmetry, where $\mathrm{Link}(\Sigma,\gamma)$ is a topological invariant of $\Sigma$, measuring how many times it links $\gamma$. The modern way of viewing symmetries is that a symmetry \textit{is} a topological operator -- if the above correlation functions depend non-topologically on $\Sigma_{d-2}$, rather than just topological invariants like the linking number, then the symmetry is broken. This was illustrated in Figure~\ref{fig:breaking}, for example.

An additional difference between the two gauge theories is that $\bR$ has no magnetic $(d-3)$-form symmetry. In $U(1)$ gauge theory there are $(d-3)$-dimensional operators known as `t Hooft defects that are the worldlines of infinitely massive probe magnetic monopoles. The possible charges of the magnetic monopoles are given by $\pi_1(U(1)) = \bZ$, so there is an `t Hooft defect for each charge. These are then the charged operators under the $U(1)^{[d-3]}$ magnetic symmetry of $U(1)$ gauge theory. In $\bR$ gauge theory however, the fact that $\pi_1(\bR)=0$ means there are \textit{no} magnetic monopoles, and so no `t Hooft defects. Therefore there is no magnetic symmetry.

Now consider adding a field of electric charge $q=1$ to $U(1)$ gauge theory. For simplicity, let this field be a complex scalar. The action then becomes
\begin{equation}
    S = \int \frac{1}{2g^2} F\wedge * F + D_\mu \phi^\dagger D^\mu \phi
\end{equation}
so that the current is no longer conserved:
\begin{equation}
    d*F \neq 0 \,.
\end{equation}
This means the operator $\cU_\alpha (\Sigma_{d-2})$ is no longer topological! Again, we can see this as the result of the Wilson lines all becoming endable. If we add a charge $N$ particle, then the symmetry will be broken to a $\bZ_N^{[1]}$ that measures the charge of the Wilson line modulo $N$.

This screening argument works very nicely for $U(1)$ gauge theory. However, for $\bR$ gauge theory things are trickier. The charges under a $U(1)$ gauge symmetry are quantised to be $\bZ$ -- the unit charge $q=1$ generates all of $\bZ$, so that all Wilson lines become endable by adding a single field. The charges under an $\bR$ gauge symmetry are not quantised, with $q\in\bR$. Then, if we add a charge-$q$ field to screen the charge $q$ Wilson line, there will still be a continuum of unscreened Wilson lines.
In fact, adding \emph{any} countable set of fields $\phi_q$ will leave uncountably many Wilson lines unscreened. Hence to break the $\bR^{[1]}$ symmetry, we need to add uncountably infinitely many fields to the EFT.

 So, we can see that non-compact gauge groups are forbidden in quantum gravity because we seemingly cannot break the global symmetries to satisfy the No Global Symmetries conjecture. The only way to do so would be to add uncountably infinitely many scalars to have all Wilson lines end. So why can't we just gauge the $\bR$ global symmetry? The reason is that this produces another global $\bR$ symmetry. To see this, let us gauge $\bR^{[1]}$ by introducing a 2-form $\bR$ gauge field $B$ and then coupling it to the current of the centre symmetry, $*F$, in a gauge-invariant way:
\begin{equation}
    S = \int \frac{1}{2g^2}(F-B)\wedge*(F-B) \,.
\end{equation}
Why is this how we gauge? One can see the action of the 1-form centre symmetry as $A \rightarrow A + \alpha$ where $\alpha$ is a constant 1-form valued in $\bR$. This leaves $F$ invariant, so that $F\wedge*F$ is also invariant. Gauging the symmetry means letting $\alpha$ depend on spacetime, so that now
\begin{equation}
    F=dA \rightarrow dA + d\alpha(x).
\end{equation}
This means the kinetic term $F\wedge*F$ is not gauge invariant, and the gauge field $B$ is there to cancel the gauge variance by transforming as
\begin{equation}
    B \rightarrow B + d\alpha.
\end{equation}
This is really the purpose of a gauge field in life, to make an action gauge invariant after letting symmetry parameters depend on spacetime.

Now that we have introduced the gauge field to the action, we have a choice: should we let $B$ be flat, i.e. $dB=0$, or dynamical, $dB \neq0$? Both are valid, and are two different ways of gauging a continuous symmetry, which are sometimes referred to as flat and dynamical gauging, respectively. If we let $B$ be flat, then we can see that there are topological Wilson lines of $B$,
\begin{equation}
    \Tilde{\cW}_{\Tilde{q}}(\Sigma_2) = e^{i\Tilde{q}\int_\Sigma B},
\end{equation}
which generate an $\widehat{\bR}=\bR$ $(d-3)$-form symmetry. This is the common statement that flat gauging a $G^{[p]}$ symmetry gives a dual $\widehat{G}^{[d-p-2]}$ symmetry, where $\widehat{G}\coloneqq \Hom(G,U(1))$ is the Pontryagin dual of $G$. %
\footnote{
    In this work we consider Wilson lines labelled by the Pontryagin dual of the gauge group throughout. The Pontryagin dual of $G$ corresponds to finite-dimensional unitary representations. Notably, $G=\bR$ has non-unitary representations labelled by $\bC$, of which $\widehat{G}=\bR$ are a subset. As such, we do not consider these non-unitary representations in this work.
} %
Importantly, $\bR$ is self-Pontryagin dual, so gauging an $\bR$ global symmetry in this way just reproduces another $\bR$ global symmetry of a different degree.

Instead of flat gauging, we can let $B$ be a dynamical gauge field and add a kinetic term to the action:
\begin{equation}
    S = \int \frac{1}{2g^2}(F-B)\wedge*(F-B) + \frac{1}{2h^2}H\wedge* H \,,
\end{equation}
where locally $H=dB$. Then, we will have a Bianchi identity $dH=0$, which means
\begin{equation}
    e^{i\alpha\int H}
\end{equation}
is topological, with $\alpha \in \bR$. This means that we have a dual $\bR^{[d-4]}$ symmetry coming from dynamical gauging. This is just the statement that dynamically gauging an abelian $G^{[p]}$ symmetry produces a dual $G^{[d-p-3]}$ symmetry.

Either way we choose to gauge $\bR^{[1]}$, we get a dual $\bR$ symmetry of a different degree taking its place. One potential way out is to instead gauge $2\pi\bZ^{[1]} \subset \bR^{[1]}$. As $\bZ$ is discrete, the only option is to flat gauge it. That means we are considering the action
\begin{equation}
    S = \int \frac{1}{2g^2}(F-B)\wedge*(F-B) \,,
\end{equation}
where $F \rightarrow F +2\pi dn$ under $A \rightarrow A + 2\pi n$, where we have let $\alpha = 2\pi n$ depend on spacetime, where $\int_\gamma n = n(\gamma)$ is an integer. This is essentially just forcing $A$ to be subject to large gauge transformations. The transformation $F \rightarrow F + 2\pi dn$ is how $U(1)$ gauge field strengths (or rather, their corresponding Chern classes) transform under large gauge transformations in the Hopkins-Singer formalism of $U(1)$ gauge theories \cite{Hopkins:2002rd}, and this $B$ field is here to cancel out this variance. This is similar to the modified Villain formulation of $U(1)$ lattice gauge theory \cite{Sulejmanpasic:2019ytl,Gorantla:2021svj}.

Now, the only gauge invariant Wilson lines are $\cW_q$ with $q\in \bZ$, which we could explicitly break by adding a charge-1 scalar/fermion. However, the gauge group has become $\bR / 2\pi \bZ \cong U(1)$. That is to say, this option of gauging $2\pi \bZ^{[1]}\subset \bR^{[1]}$ sidesteps the issue -- we no longer have a non-compact gauge group, so there is no issue with breaking it explicitly.

So, it seems that unless we add uncountably infinitely many fields to screen $\cW_q$ for all $q\in \bR$, there will always be a generalised global symmetry associated to an $\bR$ gauge symmetry. What we wish to do for the remainder of this work is to see that, while adding infinitely many fields to a $d$-dimensional EFT is bad news from the $d$-dimensional perspective, that for certain restricted regions of parameter space we get a very reasonable $(d+1)$-dimensional EFT. That is, we can see such $\bR$ gauge theories with no global symmetries, coming from a complete spectrum of operators, as a decompactification limit of a $(d+1)$-dimensional theory on $\cM_d \times \bR$. Before doing this in section §\ref{sec:extra-dimensions}, we will first show that adding uncountably infinitely many fields allows us to break the 0-form symmetries of the theory.

\section{Breaking symmetries with an infinite number of fields} \label{sec:breaking-symmetries}

	In \cite{Banks:2010zn} it was argued that continuous gauge groups could not be compact in theories of quantum gravity.
	The argument went that if there was an $\bR$ gauge symmetry, then we could have local operators of relatively irrational charge, say $1$ and $\sqrt2$.
	Then:
	\begin{enumerate}
		\item There would be a U(1) global symmetry under which the operator of gauge charge $q=\sqrt2$ has global charge $Q=1$, whilst the operator of gauge charge $q=1$ has global charge $Q=0$, contradicting the no global symmetries conjecture.
		\item The operators could be used to construct a large number of states, so as to violate the covariant entropy bound (CEB).
	\end{enumerate}
	Focusing on the first argument for now, in the theory with only fundamental fields of charge 1 and $\sqrt2$ there is no gauge invariant (perturbative) interaction that explicitly breaks the above global symmetry. However we needn't restrict ourselves to just these charges -- any $q\in\bR$ is a possible charge in the theory. Could it be possible that by adding more charges to the theory, we could eventually explicitly break the global symmetry?

	As we will see in this section, it turns out that if we add only finitely many fields, then the answer is no, as suggested in \cite{Banks:2010zn}. However, by adding uncountably infinitely many fields, we can in fact explicitly break the symmetry, albeit in a somewhat subtle way -- in §\ref{sec:noether-current-integrability} we will show that once we add uncountably infinitely many fields with all possible charges $q\in\bR$, the Noether current for the symmetry becomes ill-defined, and so then does the topological operator generating the symmetry. Of course, adding infinitely many fields does not constitute a valid EFT in $d$-dimensions. However, in §\ref{sec:extra-dimensions} we will see that, for certain regions of parameter space, adding these fields produces a valid $(d+1)$-dimensional EFT. That is, we explicitly break all global symmetries of an $\bR$ gauge theory, but the theory decompactifies into an extra dimension!

\subsection{Global 0-form symmetries from a finite number of fields}\label{sec:flavour-symmetry-for-finitely-many-fields}
	Consider a complex scalar field $\phi_q$ of gauge charge $q$, for some values of $q$, with $\phi_q^\dagger := \phi_{-q}$, such that the fields have charge $Q_q$ under a global 0-form symmetry.
	That is, the global symmetry is
	\begin{equation}
		\phi_q \mapsto e^{i\alpha Q_q}\phi_q\,,
	\end{equation}
	where $\alpha\in\bR$ parametrises the global symmetry.

	Let us consider the above example with both $q=1,\sqrt2$, and assume we have a global symmetry with $Q_{\pm1}=0$, $Q_{\pm\sqrt2}=\pm1$.
	No interactions can explicitly break this symmetry, since gauge invariance imposes that the net number of copies of the charge $\sqrt2$ field%
    \footnote{
        Where by net number of the charge $q$ field we mean the number of copies of $\phi_q$ minus the number of copies of $\phi_{-q}$.
    } %
    is $0$, and the net number of copies of the charge $1$ field is $0$.
    Hence every interaction also preserves this global symmetry.

	But if we were to also add a field of gauge charge $q=2\sqrt2$, then we could create interactions of the form
	\begin{equation}
		\phi_{\sqrt2}\phi_{\sqrt2}\phi_{2\sqrt2}^\dagger\,.
	\end{equation}
	This has $2$ net number of $\sqrt2$ fields, so if we take the $\phi_{2\sqrt2}$ field to be uncharged, the global symmetry would be broken. But since we modified the theory by adding a field, we need to consider \emph{all} possible $Q_q$ charges of the fields to check if a global symmetry persists. Indeed, if we assign global charge $Q_{2\sqrt2}=+2$ to $\phi_{2\sqrt2}$, the global symmetry is in fact preserved. Therefore, adding fields (even infinitely many) of charges in $\operatorname{span}_\bQ\brap{1,\sqrt2}$ doesn't break the global $\bR$ symmetry, by similar reasoning.

	Suppose then we were to add a charge not in that span, say $\phi_\pi$.
	In order for us to try to break the original global symmetry via interactions, we need to add yet another field to ensure gauge invariance, for example $\phi_{\brap{\pi+\sqrt2}}$.
	Then we can add interactions of the form
	\begin{equation}
		\phi_{\sqrt2}\phi_{\pi}\phi_{\brap{\pi+\sqrt2}}^\dagger\,.
	\end{equation}
	Here the situation is even worse! Not only is there an assignment of global charge that preserves the global symmetry, there is actually an infinite choice --
	we are free to pick the charge $Q_\pi$ of $\phi_\pi$ to be \emph{any} real number, say $\cQ$. Then as long as we pick the charge of $\phi_{\brap{\pi+\sqrt2}}$ to be $\cQ+1$, the symmetry is preserved.
	Therefore we have in fact enhanced the global symmetry to an $\bR^2$ symmetry, parametrised by $\alpha$, $\beta$, and given by:
	\begin{align}
		\phi_1 &\mapsto \phi_1 \,,
		\\
		\phi_{\sqrt{2}} &\mapsto e^{i\alpha} \phi_{\sqrt{2}}\,,
		\\
		\phi_{\pi} &\mapsto e^{i\beta} \phi_{\pi}\,,
		\\
		\phi_{\brap{\pi+\sqrt2}} &\mapsto e^{i\brap{\alpha+\beta}} \phi_{\brap{\pi+\sqrt2}}
		\,.
	\end{align}
	By following similar reasoning, we find that adding any finite number of fields preserves some global symmetry, and could in fact enhance the global symmetry. Of course, if we add only finitely many fields then there will also exist Wilson lines that cannot be screened, so there will be some unbroken 1-form symmetry as well as the 0-form symmetries.

\subsection{Adding infinitely many fields}\label{sec:constraining-Q-by-interactions}
	It turns out that by adding uncountably infinitely many fields, we can break all global symmetries. Indeed, there are uncountably infinitely many Wilson lines to screen, so we need uncountably infinitely many local operators to end them. We therefore assume that we have a field $\phi_q$ for \emph{every} $q\in\bR$, with $\phi_{q}^\dagger=\phi_{-q}$. %
    \footnote{
        See Appendix \ref{appendix:finite-charge-interval} for an alternative case where we instead assume a field $\phi_q$ for {every} $q\in\brap{-L,L}$, for finite $L$, with $\phi_{q}^\dagger=\phi_{-q}$.
    } %
	We shall show that the global 0-form symmetry can now be broken by interactions.

	Suppose that a global symmetry remains after adding interaction terms, with global charges $Q_q$.
	The only gauge invariant interactions we can add are ones in which the sum of the gauge charges of the interacting particles is zero:
	\begin{equation}
		c_{q_1,q_2,\dots,q_n}\phi_{q_1} \phi_{q_2} \dots \phi_{q_n} \delta_{\sum_{i=1}^n q_n,\ 0}\,,
	\end{equation}
	where $c_{q_1,q_2,\dots,q_n}$ is a coupling constant,
	and $\delta_{\sum_{i=1}^n q_n,\ 0}$ ensures gauge invariance of the coupling.

	Under the global transformation, the above term transforms to
	\begin{equation}
		e^{i\alpha\brap{\sum_{i=1}^{n}Q_{q_i}}} \cdot c_{q_1,q_2,\dots,q_n}\phi_{q_1} \phi_{q_2} \dots \phi_{q_n} \cdot \delta_{\sum_{i=1}^n q_n,\ 0}\,.
	\end{equation}
	Hence, the global symmetry remains unbroken if and only if
	\begin{equation}
		{\sum_{i=1}^{n}Q_{q_i}}=0 \,,
	\end{equation}
    where $\sum_{i=1}^{n}{q_i}=0$.

	Let's assume that we have an interaction of this form for all fields $\phi_{q_1},\phi_{q_2},\dots,\phi_{q_n}$ that satisfy $\sum_{i=1}^{n} q_i=0$.
	How does this restrict the possible global charges $Q_q$?
    Enumerating $n$:
	\begin{itemize}
		\item All $n=2$ interactions are of the form $\phi_q\phi_{-q}$. This imposes that for all $q$:
		\begin{equation}\label{eq:Q-condition-negation}
			Q_{-q}=-Q_{q}\,.
		\end{equation}
		\item $n=3$ interactions impose
		\begin{equation}
			Q_{q_1}+Q_{q_2} = -Q_{-\brap{q_1+q_2}}\,,
		\end{equation}
		for all $q_1$, $q_2$. Combining with the $n=2$ restriction, we find that
		\begin{equation} \label{eq:Q-condition-addition}
			Q_{\brap{q_1+q_2}} = Q_{q_1}+Q_{q_2} \,.
		\end{equation}
		\item $n>3$ interactions impose
		\begin{equation}
			Q_{q_1}+Q_{q_2}+\dots+Q_{q_{n-1}} = -Q_{-\brap{q_1+q_2+\dots+q_{n-1}}}\,,
		\end{equation}
		for all $q_1, q_2, \dots q_{n-1}$.
		But this already follows from imposing the $n=2$ and $n=3$ conditions.
	\end{itemize}
	So adding interactions beyond the cubic ones imposes no further restrictions on $Q_q$.
	Further, if we assume \eqref{eq:Q-condition-addition}, then for $m\in\bN$,
	\begin{equation}
		Q_{mq} = mQ_{q}\,,
	\end{equation}
	and by using \eqref{eq:Q-condition-negation}, we can promote this to being true for all $m\in\bZ$.
	Then, using this repeatedly, we find for $a,b\in\bZ$, $b\neq0$, that
	\begin{equation}
        Q_{\frac{a}{b}q} = aQ_{\frac{1}{b}q}
        \,,\;\;\;\text{and}\;\;\;
        Q_q=Q_{b\frac{1}{b}q} = bQ_{\frac{1}{b}q}
        \,,
	\end{equation}
	so
	\begin{equation}
		Q_{\frac{a}{b}q} = \frac{a}{b}Q_q\,.
	\end{equation}

	So, by including all gauge invariant quadratic and cubic terms, we can constrain the $Q_q$ to obey the following condition:
	\begin{align}\label{eq:restrictions-on-Q}
		\boxed{Q_{aq_1+bq_2} = aQ_{q_1} + bQ_{q_2}}
	\end{align}
	for all $q_1,q_2 \in \bR$, and all $a,b\in\bQ$.

    Then, the most general $Q_q$ is constructed as follows:
	\begin{enumerate}
		\item Pick a basis $\cB$ of $\bR$ over $\bQ$. %
            \footnote{
                The existence of such a basis follows from the axiom of choice \cite{MathWorld:hamel:basis,nlab:basis_of_a_vector_space}, which we assume here.
            } %
            That is, each $q\in\bR$ can be written uniquely as some $\sum_{r \in B\brap{q}} a_r r$, for some $a_r\in\bQ$, and $B\brap{q}$ a \emph{finite} subset of $\cB$.
		\item To each element in $r\in\cB$, choose a value $\lambda_r \in \bR$.
		\item Then for $q=\sum_{r \in B\brap{q}} a_r r$, we define $Q_q = \sum_{r \in B\brap{q}} a_r \lambda_r$.
	\end{enumerate}
	For example:
	\begin{itemize}
		\item If we take $\lambda_r = 0$ for all $r\in \cB$, then $Q_q\equiv0$, i.e. the global symmetry is trivial.
		\item If we take $\lambda_r=r$, then $Q_q=q$, i.e. the global symmetry is the rigid part of the gauge symmetry, so the global symmetry is trivial.
		\item If we take the basis $\cB$ to include $1$ and $\sqrt2$, then any such $\cB$ and $\lambda$ such that $\lambda_1=0$ and $\lambda_{\sqrt2}=1$ is a global symmetry as in the previous example.
	\end{itemize}
	This seems problematic -- the global symmetry has appeared to persist.

	However, while this is an operation on the Lagrangian that leaves the action unchanged, it turns out that there is no corresponding topological operator, and hence this is not a global symmetry in the modern sense. We demonstrate this in the next subsection.

\subsection{Infinite number of fields and the failure of the Noether current} \label{sec:noether-current-integrability}

	We can write the Lagrangian of $\bR$ gauge theory with the infinitely many fields added as
    \begin{align} \label{eq:lagrangian}
		\cL &= \frac{1}{2g^2}F_{\mu\nu}F^{\mu\nu}  \\
        &-\int \dbar q \cdot \left[ g^{\mu\nu}\brap{D_\mu\phi_q\brap{x}}^\dagger\brap{D_\nu\phi_q\brap{x}} +m_q^2 \phi_q^\dagger \phi_q + \int \dbar q' \cdot c_{q,q'} \phi_q \phi_{q'} \phi_{-(q+q')} + \mathrm{h.c.} \right]\,, \nonumber
	\end{align}
	where
	\begin{equation}
		\dbar^d p := \frac{d^d p}{\brap{2\pi}^d}\,,
	\end{equation}
	and
	\begin{equation}
		D_\mu\phi_q = \partial_\mu \phi_q - iqA_\mu \phi_q\,.
	\end{equation}
    We have assumed that the fields are scalars for simplicity. The $c_{q,q'}$ parameter is the coupling constant between these scalars.

	The Noether current for the global 0-form symmetry is
	\begin{equation}\label{eq:non-evaluateable-noether-current}
		J_\mu = \int \dbar q Q_q J^q_\mu\,,
	\end{equation}
	where $J^q_\mu=i\bras{\phi_q^\dagger D_\mu\phi_q - \phi_q D_\mu\phi_q^\dagger}$.
	From this, one defines the topological operator generating the symmetry as
	\begin{equation}\label{eq:candidate-topological-operator}
		U_\alpha\brap{\Sigma_{d-1}} = \exp\brap{i\alpha\int_{\Sigma_{d-1}}\star J}\,.
	\end{equation}
	However, to define the topological operator in this way, we must be able to evaluate the integral. In particular, we must be able to evaluate \eqref{eq:non-evaluateable-noether-current}. This requires $Q_q$ to be integrable.

	When is $Q_q$ integrable%
    \footnote{
        In this section we handle the case of Riemann integrability; for comments on trying to evaluate the integral via some other sort of integration method, see Appendix \ref{appendix:other-methods-of-integration}.
    }%
    ?
    To answer this question, we shall make use of the fact that if two Riemann integrable functions agree on a dense subset%
    \footnote{
        A subset $D$ of $S$ is said to be dense in $S$ if for any open subset $U$ of $S$, there is an element of $D$ inside $U$.
    } %
    of an interval $I\subset\bR$, then their Riemann integrals agree over that same interval.

    With this in mind, assume $Q_q$ is Riemann integrable.
    We then pick any non-zero $q'\in\bR$, and define $k:=\frac{Q_{q'}}{q'}$.
    Then, using \eqref{eq:restrictions-on-Q}, we have $Q_q = kq$ for all $q$ in $S_{q'}:=\operatorname{span}_\bQ\brap{q'}$.
    $S_{q'}$ is dense in $\bR$, and the function $f\brap{q}=kq$ is Riemann integrable as it is a continuous function. So, as $Q_q$ is Riemann integrable, its integral is equal to that of $f\brap{q}$.

    We could similarly pick any non-zero $\Tilde{q}\in\bR$ and define $\Tilde{k}:=\frac{Q_{\Tilde{q}}}{\Tilde{q}}$.
    Following similar reasoning, we also conclude that the Riemann integral of $Q_q$ is equal to that of $\Tilde{f}\brap{q}=\Tilde{k}q$.

    But then the Riemann integrals of $f\brap{q}$ and $\Tilde{f}\brap{q}$ must also agree, imposing $k=\Tilde{k}$.
    Feeding this back into the definitions of $k$ and $\Tilde{k}$ and rearranging, we find $Q_{\Tilde{q}}=k\Tilde{q}$.
    Non-zero $\Tilde{q}$ was arbitrary, so $Q_{\Tilde{q}}=k\Tilde{q}$ for all non-zero $\Tilde{q}\in\bR$, and it also holds for $\Tilde{q}=0$ (as then both sides are zero).
    Note also that the function $Q_q=kq$ is indeed Riemann integrable as it is a continuous function of $q$.

    Hence the most general Riemann integrable function $Q_q$ consistent with \eqref{eq:restrictions-on-Q} is
    \begin{equation}
        Q_q = kq
        \,,
    \end{equation}
    for some constant $k\in\bR$.

    If $k=0$, then we do not have a global symmetry, since the resulting Noether current and topological operators are trivial.
	If $k\neq0$, then we find that $J$ is $k$ times the Noether current associated with conservation of gauge charge, and so \eqref{eq:candidate-topological-operator} generates gauge transformations with parameter $k\alpha$, so does not generate a global symmetry. An alternative way of viewing this is that the global symmetry transformation can be undone by a gauge transformation, and so all fields in the same orbit of the global symmetry are also in the same gauge orbit, and therefore physically equivalent. We cannot define a topological operator for any other choice of $Q_q$, so therefore the global symmetry is broken!

	So from the modern perspective, we have successfully broken both the global 1-form and 0-form symmetries by adding uncountably infinitely many fields, as the Wilson lines are all endable and the topological operator generating the would-be 0-form symmetry cannot be defined.

    We started this section by recapping the reasons why non-compact gauge theories are in the swampland. The one we have considered in this section, and will continue to consider for the remainder of this work, was that such gauge symmetries always seem to be accompanied by unbreakable global symmetries. We considered adding uncountably infinitely many scalar fields so as to break these symmetries. However, there existed a second reason for the absence of non-compact gauge symmetries, namely the violation of the covariant entropy bound (CEB). Having a continuum of fields sounds like a disaster for the CEB in $d$-dimensions. However, what we will argue in the next section is that, for certain regions of parameter space, the charges $q\in \bR$ are actually just momenta in an extra non-compact direction. Fourier transforming these fields back to position space in this extra direction produces a $(d+1)$-dimensional theory with a single scalar field in the presence of $(d+1)$-dimensional gravity, without any gauge symmetries. Such a $(d+1)$-dimensional theory, as it is seemingly quite mundane in $d+1$ dimensions, would be expected to satisfy the $d+1$-dimensional CEB, so that we avoid the second argument for having no non-compact gauge symmetry in the original $d$-dimensions. The decompactification procedure will be explained in more detail in the following section.

\section{Decompactification limits} \label{sec:extra-dimensions}
    In the previous section, we have shown how adding uncountably infinitely many fields $\phi_q$ breaks both the 1-form $\bR$-symmetry and the potential 0-form symmetries that could occur when adding only finitely many $\phi_q$.
	But does it make sense to talk about a theory with uncountably many fields?
	Indeed, if we add even \textit{countably} infinitely many light fields like in a Kaluza-Klein reduction on $S^1(R)$ then the species scale, the UV cut-off scale at which quantum gravitational effects must be included, would be pushed down to zero as $R\rightarrow \infty$. In this limit, the EFT will no longer make sense at any energy level \cite{Arkani-Hamed:2005zuc,Distler:2005hi,Dimopoulos:2005ac,Dvali:2007hz,Dvali:2007wp,Castellano:2022bvr,Castellano:2023stg,Castellano:2023jjt}:
    \begin{equation}
        \Lambda_\mathrm{sp} \approx \frac{M_\mathrm{Pl;d}}{N^{\frac{1}{d-2}}} \xrightarrow{N\rightarrow \infty} 0 \,,
    \end{equation}
    where $N$ is the number of light particles, or `species', whose mass is smaller than $\Lambda_\mathrm{sp}$, and $M_\mathrm{Pl;d}$ is the $d$-dimensional Planck scale.

	We propose that a way to make sense of an EFT with uncountably many fields is to consider it as a theory in one higher dimension, akin to a Kaluza-Klein reduction on $\bR$, rather than a compact direction like $S^1$. Consider coupling the Lagrangian from \eqref{eq:lagrangian} to gravity, where we continue to assume the infinite fields are massive complex scalars for simplicity. We then have an action
    \begin{align} \label{eq:4d-lagrangian}
        &S_d = \int d^dx \sqrt{-g} \cdot \Bigg\{ \frac{1}{G_N^{(d)}}\cR^{(d)} + \frac{1}{2g^2}F_{\mu\nu}F^{\mu\nu} \\
        &-\int \dbar q \cdot \left[ g^{\mu\nu}\brap{D_\mu\phi_q\brap{x}}^\dagger\brap{D_\nu\phi_q\brap{x}} +m_q^2 |\phi_q|^2 + \int \dbar q' \cdot c_{q,q'} \cdot \phi_q \phi_{q'} \phi_{-(q+q')} + \mathrm{h.c.} \right] \Bigg\}\,, \nonumber
    \end{align}
    where $G_N^{(d)} = M_\mathrm{Pl;d}^{2-d}$, with $M_\mathrm{Pl;d}$ the $d$-dimensional Planck scale. What we aim to show is that for some specific parameters this action is equivalent to a $D$-dimensional action, where $D=d+1$. Not all EFTs will admit such a decompactification, and the aim of this section is to determine which submanifold of the parameter space of this theory is valid.

    \subsection{A Kaluza-Klein reduction on \texorpdfstring{$\bR$}{R}} \label{sec:KK-reduction}
    Consider the $D$-dimensional action of a single massive complex scalar field coupled to gravity
    \begin{equation} \label{eq:Einstein-Scalar}
        S_D = \int_{\cM_d \times X} d^dx\cdot dy \cdot \sqrt{-G} \cdot \brac{\frac{1}{G_N^{(D)}}\cR^{(D)} + G^{MN}\brap{\partial_M\Phi}^\dagger\brap{\partial_N\Phi} +M^2 |\Phi|^2}\,,
    \end{equation}
    where we have chosen a background spacetime of the form $\cM_d \times X$, with metric $G_{MN}$, such that $M,N = 0,\dots,D-1$. Let us consider the usual Kaluza-Klein reduction on $X = S^1(R)$. We can take the field $\Phi(x,y)$ where $y\in S^1(R)$, and make a Fourier expansion
    \begin{equation}
        \Phi(x,y) = \sum_{n\in \bZ} \phi_n(x) e^{\frac{iny}{R}}
    \end{equation}
    so that $\Phi(x,y+2\pi R) = \Phi(x,y)$. Here, $n$ is the momentum of $\Phi$ in the circular direction, which is naturally quantised. If we then dimensionally reduce on this circle, we get $d$-dimensional Fourier modes $\phi_n(x)$ in the lower dimension with masses $m_n^2 = M^2 +\frac{n^2}{R^2}$.
    We can write the Kaluza-Klein ansatz for the metric as
    \begin{equation}
        G_{MN}\brap{X} =
		\begin{pmatrix}
			g_{\mu\nu}\brap{x} + \frac1{\lambda}A_\mu\brap{x}A_\nu\brap{x} & \frac1{\lambda}A_\mu\brap{x}\\
			\frac1{\lambda}A_\nu\brap{x} & \frac1{\lambda}
		\end{pmatrix} \label{eq:D-dim-metric} \,,
    \end{equation}
    where $\lambda$ is usually called the dilaton field. Dimensionally reducing the $D$-dimensional Einstein-Hilbert term reproduces the familiar $d$-dimensional Einstein-Maxwell-Dilaton theory
    \begin{equation} \label{eq:Einstein-Maxwell-Dilaton}
        \frac{1}{G_N^{(d)}}\int_{\cM_d} d^dx \cdot\sqrt{-g} \cdot \brac{ \cR^{(d)} + \frac{1}{2\lambda}F_{\mu\nu}F^{\mu\nu} + \frac{1}{\lambda^2} \partial_\mu \lambda \partial^\mu \lambda} \,,
    \end{equation}
    where $\mu = 0,\dots,d-1$. The large diffeomorphisms of $S^1$, i.e. the shifts in winding number around the circle, impose large gauge transformations on the $d$-dimensional gauge field. As discussed in §\ref{sec:background}, the large gauge transformations make this a $U(1)$ gauge symmetry. The Fourier mode label $n$ also corresponds to the charge of $\phi_n$ under this gauge symmetry, interpreted as the winding mode of the $\Phi$ field around $S^1(R)$.

    If we instead consider $X = \bR$, then the $\Phi(x,y)$ field will no longer have quantised momentum in this direction. That is, the Fourier expansion does not need to ensure periodicity $\Phi(x,y)=\Phi(x,y+2\pi R)$, and so we instead obtain an ordinary Fourier transform
    \begin{equation} \label{eq:fourier-transform}
        \Phi(x,y) = \int \dbar q \cdot \phi_q(x) \cdot e^{iqy} \,,
    \end{equation}
    where $\Tilde{\phi}(x,q)$ is the Fourier mode of $\Phi(x,y)$ with momentum $q$ in the $X=\bR$ direction. If we consider the form of the $D$-dimensional inverse metric,
    \begin{equation}
        G^{MN}\brap{x,y} =
		\begin{pmatrix}
			g^{\mu\nu}\brap{x} & -g^{\mu\nu}\brap{x}A_\nu\brap{x}\\
			-A_\mu\brap{x}g^{\mu\nu}\brap{x} & A_\mu\brap{x}g^{\mu\nu}\brap{x}A_\nu\brap{x} + \lambda(x)
		\end{pmatrix}\,, \label{eq:D-dim-inverse-metric}
    \end{equation}
    then we can consider the dimensional reduction of the kinetic term of $\Phi$ after substituting in \eqref{eq:fourier-transform}. After some computation, one derives%
    \footnote{
        Strictly speaking we get an extra overall factor of $\sqrt{\lambda^{-1}}$, coming from $\sqrt{-G} = \sqrt{\lambda^{-1}} \sqrt{-g}$. Note also that, from the definitions above, not all the $d$-dimensional quantities have the correct mass dimensions in $d$-dimensions. Appendix \ref{appendix:matching-mass-dimensions} details how to remedy these issues by redefining various quantities, but we avoid writing such re-definitions in the main text so as to avoid clutter.
    }
    \begin{equation} \label{eq:scalar-decompactification}
        \int_X dy \cdot \sqrt{-G} \cdot  G^{MN}\brap{\partial_M\Phi}^\dagger\brap{\partial_N\Phi} = \int \dbar q \cdot \sqrt{-g} \cdot \brac{ g^{\mu\nu}\brap{D_\mu\phi_q\brap{x}}^\dagger\brap{D_\nu\phi_q\brap{x}} + \lambda q^2 |\phi_q|^2 } \,.
    \end{equation}
    Computing also
    \begin{equation}
        \int_X dy \cdot \sqrt{-G} \cdot M^2 |\Phi(x,y)|^2 = \int \dbar q \cdot \sqrt{-g} \cdot M^2 |\phi_q|^2 \,,
    \end{equation}
    we see that the $D$-dimensional scalar sector dimensionally reduced on $X=\bR$ produces a $d$-dimensional theory with infinitely many scalar fields, but only when the mass spectrum is of the form
    \begin{equation} \label{eq:mass-q}
        m_q^2 = M^2 + \lambda q^2\,.
    \end{equation}

    One can show, by using
    \begin{equation}
        \partial_y \Phi(x,y) = \int \dbar q \cdot (iq) \cdot \phi_q(x) \cdot e^{iqy} \,,
    \end{equation}
    that gauge invariant derivative interactions in the $D$-dimensional theory allow for a mass spectrum
    \begin{equation}\label{eq:mass-q-non-zero-cn}
        m_q^2 = M^2 + \lambda q^2 + \sum_{n\geq 2} c_n q^{2n} \,,
    \end{equation}
    though we will see in §\ref{sec:extra-dim-gauge-kinetic-term} that setting $c_n=0$ for all $n$ is required for the $d$-dimensional EFT to satisfy additional swampland constraints. Note also that these higher-derivative interactions in $D$-dimensions are irrelevant in $D$ dimensions.

    We can also reproduce the $d$-dimensional cubic interaction terms from a $D$-dimensional cubic interaction. One computes, with $c_\Phi$ a coupling constant,
    \begin{equation}
        \int dy \cdot c_\Phi \Phi \Phi \Phi + \mathrm{h.c.} = \int \dbar q \cdot \dbar q' \cdot c_\Phi \phi_q \phi_{q'} \phi_{-(q+q')} + \mathrm{h.c.}
    \end{equation}
    which poses another constraint on the ability for the $d$-dimensional theory to decompactify to $D$-dimensions:
    \begin{equation}\label{eq:constraint-on-3-point-interaction}
        c_{q,q'} = c_{\Phi}
    \end{equation}
    for all $q,q'$. %
    \footnote{\label{footnote:more-general-cubic-interaction}
        If we allowed higher derivative terms in $D$-dimensions, we could get a $c_{q,q'}$ that depends upon $q$ and $q'$. The corresponding $D$-dimensional operators are irrelevant in $D$-dimensions, and whether $c_{q,q'}$ is constant in $q$, $q'$ doesn't affect our discussions except for in our discussion of $-1$-form symmetries in \S\ref{sec:extra-dim-gauge-kinetic-term}.
    }

    Therefore, while not all $d$-dimensional EFTs of the form \eqref{eq:4d-lagrangian} can decompactify, if we restrict to the following submanifold of parameter space:
    \begin{equation}
        m_q^2 = M^2 + \lambda q^2,\quad c_{q,q'} = c_{\Phi}\,,
    \end{equation}
    then the EFT can decompactify to a theory with just a single scalar coupled to gravity in $D$-dimensions. In fact, one can obtain additional constraints on the parameter space by considering the $d$-dimensional Einstein-Maxwell-Dilaton sector. We will do this in the following subsection.

\subsection{The gauge kinetic term, the Weak Gravity Conjecture, and free parameters}\label{sec:extra-dim-gauge-kinetic-term}

    So far we have only considered constraints on the $d$-dimensional EFT from the scalar sector. We now turn to the Einstein-Maxwell-Dilaton sector. We can use the $D$-dimensional metric \eqref{eq:D-dim-metric} to write a $D$-dimensional Einstein-Hilbert term as a $d$-dimensional Einstein-Maxwell-Dilaton term, as in \eqref{eq:Einstein-Maxwell-Dilaton},
    \begin{equation*}
        \frac{1}{G_N^{(d)}}\int_{\cM_d} d^dx \cdot\sqrt{-g} \cdot \brac{ \cR^{(d)} + \frac{1}{2\lambda}F_{\mu\nu}F^{\mu\nu} + \frac{1}{\lambda^2} \partial_\mu \lambda \partial^\mu \lambda}\,,
    \end{equation*}
    which one can compare with the generic Einstein-Maxwell term of our $d$-dimensional EFT in \eqref{eq:4d-lagrangian}:
    \begin{equation}\label{eq:Einstein-Maxwell-descended-to-d-dimensions-no-lambda-kinetic-term}
        \int d^dx \sqrt{-g} \cdot \brac{ \frac{1}{G_N^{(d)}}\cR^{(d)} + \frac{1}{2g^2}F_{\mu\nu}F^{\mu\nu} } \,.
    \end{equation}
    A main difference is that there is no kinetic term for $\lambda$ in the latter -- we will discuss this momentarily. First, we can see that for the EFT to decompactify we must impose a constraint on $g^2$:
    \begin{equation}\label{eq:g-vs-lambda}
        M_\mathrm{Pl;d}^{d-2}\cdot g^2 = \lambda\,.
    \end{equation}
    Theories without this constraint have no such embedding into our $D$-dimensional metric in \eqref{eq:D-dim-metric}. Imposing \eqref{eq:g-vs-lambda} has implications for other Swampland Conjectures, as we explain in the rest of this subsection.

\subsubsection{The Weak Gravity Conjecture}

    The Weak Gravity Conjecture of \cite{Arkani-Hamed:2006emk} states that there should exist at least one particle of gauge charge $q$ and mass $m$ satisfying the bound
    \begin{equation}
        m M_\mathrm{Pl;d}^{\frac{2-d}{2}} \leq q g\,. \label{eq:wgc}
    \end{equation}
    This can be phrased as the requirement that there exists at least one particle on which the force of gravity is weaker than the gauge force. A particle satisfying, but not saturating, the bound is called super-extremal, and those saturating the bound are called extremal. See \cite{Harlow:2022ich,Rudelius:2024mhq} for reviews of the Weak Gravity Conjecture.

    If we impose $M_\mathrm{Pl;d}^{d-2} \cdot g^2 = \lambda$ from \eqref{eq:g-vs-lambda}, and square both sides of the inequality, then we have that a particle satisfies the Weak Gravity Conjecture if
    \begin{equation}
        m^2 \leq  \lambda q^2\,.
    \end{equation}
    If we consider the masses of our scalars \eqref{eq:mass-q},
    \begin{equation*}
        m_q^2 = M^2 + \lambda q^2
        \,,
    \end{equation*}
    where $M^2$ is the mass of the $D$-dimensional scalar, then we see that to satisfy the Weak Gravity Conjecture we require $M^2=0$. This was the case for ordinary Kaluza-Klein reductions in \cite{Heidenreich:2015nta}. All scalars will then \textit{saturate} the bound, such that we have a continuum of extremal particles.

    In §\ref{sec:KK-reduction} we we found that by including derivative interaction terms in $D$-dimensions allowed for $d$-dimensional mass terms of the form \eqref{eq:mass-q-non-zero-cn}
    \begin{equation*}
        m_q^2 = M^2 + \lambda q^2 + \sum_{n\geq 2} c_n q^{2n}
        \,.
    \end{equation*}
    With these extra terms, it is possible to satisfy the Weak Gravity Conjecture without imposing $M=0$. %
    \footnote{
        As an explicit example, consider the polynomial in $q^2$:
        \begin{equation*}
            m_q^2 = \brap{q^2-r^2}^2\brap{q^2+a^2}\,,
        \end{equation*}
        for $r,a>0$. This is manifestly non-negative, yielding real masses for all particles.
        Further, it has zeroes (only) at $q=\pm r$, and so particles of charge $q=\pm r$ satisfy \eqref{eq:wgc}.
        For any given $M>0$, $\lambda>0$, we can solve for positive $r$ and $a$ such that the above expression for $m_q^2$ is $M^2+\lambda q^2$ at leading order in $q^2$.
        Hence, for any given positive $M$ and $\lambda$, we have a $m_q^2$ of this form that satisfies the Weak Gravity Conjecture.
    }

    However, the Weak Gravity Conjecture is a conjecture about U(1) gauge fields. One of its motivations \cite{Arkani-Hamed:2006emk} is for all extremal charged black holes to be able to decay. For U(1), if there is a particle of charge $1$ that satisfies \eqref{eq:wgc}, then all extremal charged black holes may decay by repeatedly emitting copies of that particle. However, for $\bR$, there is no minimum non-zero charge. So if there is a minimum charge $q$ such that \eqref{eq:wgc} is satisfied, then there are extremal charged black holes of non-zero charge less than $q$ that can't decay.
    This suggests that the generalisation of the Weak Gravity Conjecture to $\bR$ gauge theory is that for any non-zero charge $Q$, there is at least one particle of charge less than or equal to $Q$ satisfying \eqref{eq:wgc}.

    When we impose this stronger version of the Weak Gravity Conjecture along with $M_\mathrm{Pl;d}^{d-2} \cdot g^2 = \lambda$, we find that
    we must set $M^2=0$ and all $c_n=0$.

    So in the following, we set $c_n=0$ for all $n$.

\subsubsection{Free parameters}
    Prior to imposing $M_\mathrm{Pl;d}^{d-2} \cdot g^2 = \lambda$, $g^2$ and $\lambda$ were two independent free parameters in the $d$-dimensional theory.
    It was proposed in \cite{Cordova:2019jnf,Cordova:2019uob} that free parameters can be interpreted as background gauge fields for $(-1)$-form symmetries, and as such quantum gravity should have no free parameters \cite{McNamara:2020uza,Heidenreich:2020pkc}. One can picture this argument as follows. If $g^2$ is a free parameter, then $d(F\wedge*F)=0$ due to being a top form. We can picture this as the conservation of a $0$-form current, $J_0 = *(F\wedge* F)$, $d*J_0 = 0$, and so we can define a topological $d$-dimensional operator
    \begin{equation}
        \cU(\Sigma_d) = e^{i\int F \wedge*F}
    \end{equation}
    generating a $(-1)$-form symmetry.

    The theory with $g$ and $\lambda$ both free essentially has two $(-1)$-form symmetries. By imposing $g^2 = M_\mathrm{Pl;d}^{2-d} \cdot\lambda$, we explicitly break this down to a single $(-1)$-form symmetry, that has $\lambda$ as a background gauge field.
    From the $d$-dimensional perspective, we can then gauge this symmetry by allowing $\lambda \coloneqq \lambda(x)$ to be dynamical, and adding a kinetic term%
    \footnote{
        There is no reason why one must choose the canonical kinetic term -- picking this non-canonical term allows us to embed the field into the $D$-dimensional Einstein-Hilbert term, and so this is the choice we make. One could redefine this field so that it has a canonical kinetic term, so long as the Kaluza-Klein ansatz for the metric is adjusted accordingly.
    } %
    to the $d$-dimensional action:
    \begin{equation}
        S_d \supset \int \frac{1}{\lambda^2} \partial_\mu \lambda \partial^\mu \lambda\,.
    \end{equation}
    This was the term discarded in going from \eqref{eq:Einstein-Maxwell-Dilaton} to \eqref{eq:Einstein-Maxwell-descended-to-d-dimensions-no-lambda-kinetic-term}, and so embeds naturally in the $D$-dimensional theory in the same way as for a dilaton.
    Such a field is then a modulus that controls the charge-to-mass ratio of the fields: $m_q^2 = \lambda q^2$.

    Further, the $D$-dimensional three-point interaction imposes \eqref{eq:constraint-on-3-point-interaction}%
    \footnote{
        At least assuming we restrict to relevant terms in the $D$-dimensional Lagrangian; see footnote \ref{footnote:more-general-cubic-interaction}.
    }%
    , reducing the free parameters $c_{q,q'}$ down to the single free parameter $c_\Phi$ in the $D$-dimensional theory.
    Lastly, our considerations above from the Weak Gravity Conjecture impose $M=0$, and $c_n=0$ for all $n$, removing those free parameters from the $d$-dimensional theory (and indeed also the $D$-dimensional theory).

    So we see that the restrictions on the $d$-dimensional theory imposed from being able to decompactify to $D$-dimensions, along with those from imposing the Weak Gravity Conjecture, reduce the number of free parameters down to just the single free parameter $c_\Phi$ from the $D$-dimensional theory.

\subsection{The infinite volume of \texorpdfstring{$\bR$}{R} and the vanishing of the gauge coupling} \label{sec:coupling}

    There is one final caveat to discuss. Consider a standard Kaluza-Klein reduction on some compact space $X_{D-d}$. As discussed above, the $D$-dimensional Einstein-Hilbert term becomes
    \begin{align}
        \frac1{G_N^{(D)}}\int d^DX \sqrt{-G} \cR^{\brap{D}}
        &=
        \frac1{G_N^{(d)}}\int d^dx \sqrt{-g} \cR^{\brap{d}}
        \\&-
        \frac14\int d^dx \sqrt{-g}\bras{\frac{1}{G_N^{(d)}\lambda}}F_{\mu\nu}F^{\mu\nu}
        \, + \dots,
    \end{align}
    where one defines
    \begin{equation}
        \frac1{G_N^{(d)}}
        =
        \frac1{G_N^{(D)}}\int_{X_{D-d}} d^{D-d}y
        \,.
    \end{equation}
    The integral here can be defined as the volume of $X_{D-d}$, so that we obtain
    \begin{equation} \label{eq:D-dim-Planck}
        M_\mathrm{Pl;d}^{d-2} = \mathrm{Vol}(X_{D-d}) M_\mathrm{Pl;D}^{D-2} \,,
    \end{equation}
    relating the Planck scale between the two different dimensions.

    However, in our case we are dimensionally reducing on $\bR$, so this integral will be infinitely large. Therefore, the $d$-dimensional Planck scale will also be infinitely large. Said differently, $G_N^{(d)}$ will vanish! Thankfully, the $d$-dimensional Weak Gravity Conjecture is still satisfied. However, there is another upshot of this result: the $\bR$ gauge field will now be flat, due to $g^2 = M_\mathrm{Pl;d}^{2-d} \cdot \lambda = G_N^{(d)} \lambda = 0$.

    The action of a pure gauge theory,
    \begin{equation}
        S = \int \frac{1}{2g^2} F \wedge* F \,,
    \end{equation}
    is not well-suited for discussing the $g^2=0$ limit. We can write the theory instead as \cite{Witten:1992xu}
    \begin{equation}
        S' = \int B \wedge F - \frac{g^2}{2} B \wedge * B \,,
    \end{equation}
    where $B$ is a $(d-2)$-form acting as a Lagrange multiplier. The equations of motion are $F = g^2 * B$ and $dB = 0$, so that integrating out $B$ results in the action $S$ above. Therefore, this is a completely equivalent way of describing pure gauge theory. However, notice that we can set $g^2=0$ in this formulation of the theory, giving an action
    \begin{equation}
        S'_{g^2=0} = \int B \wedge F \,.
    \end{equation}
    This is a topological theory known as a BF theory, whose equations of motion now set $F=0$. That is, the $\bR$ gauge field no longer has any propagating degrees of freedom! Therefore, as the pure $\bR$ gauge theory gets coupled to the infinity of fields that break all of its symmetries, the theory decompactifies to $D$-dimensions and we derive that $g^2 = G_N^{(d)} \cdot \lambda = 0$, so that the $\bR$ gauge field in $d$-dimensions becomes flat.

    One might hope that we could retain a non-vanishing $G_N^{(d)}$ by modifying the $d$ dimensional theory to get a finite volume for the extra dimension. However, we know of no such modification that still breaks the global symmetries that the infinity of fields were introduced to break -- for example, as discussed in Appendix \ref{appendix:finite-charge-interval}, restricting $q$ to a finite volume interval has the effect of implementing a momentum cut-off in the $D$-dimensional theory, rather than restricting its spatial volume.

    Note that $G_N^{(d)}=0$ doesn't mean we have lost the $d$-dimensional gauge symmetry: $F = dA$ is vanishing on-shell but is still invariant under the original gauge symmetry $A \rightarrow A + d\lambda$, so that this is still a symmetry of $S'_{g^2=0}$.

\section{Conclusion} \label{sec:conclusion}

Let us now summarise the main results of this work:
\begin{itemize}
    \item We discussed the original arguments for why quantum gravity cannot have non-compact gauge symmetries, namely that they are always accompanied by generalised global symmetries. We gave a proof that adding finitely many fields to the theory will never break every global symmetry. We then tried adding \textit{uncountably infinitely many fields}, one for each possible gauge charge $q \in \bR$. We showed that while naïvely a global symmetry persists, that the topological operator generating this symmetry is ill-defined. Therefore, adding a continuum of fundamental fields to the theory breaks all global symmetries in the modern sense.
    \item Having started with a theory in $d$-dimensions consisting of an $\bR$ gauge theory and a scalar field for each possible charge $q \in \bR$, we treated the charge of the field as a momentum mode in an extra dimension. By constraining the allowed mass spectrum and interaction couplings, we were able to use a Fourier transform to repackage the uncountably many $d$-dimensional fields as a single $D$-dimensional field, living in a theory in $D$ dimensions, with the metric in the higher dimension given by a combination of the lower dimensional metric, gauge field, and a modulus that controls the charge to mass ratio of the $d$-dimensional fields. The whole process is akin to undoing a Kaluza-Klein reduction on $\bR$.
    \item  Consistency requirements from the Kaluza-Klein reduction, along with the additional assumption that the $D$-dimensional scalar was massless (with no higher-derivative interactions), automatically implied the $d$-dimensional Weak Gravity Conjecture. We also briefly discussed how the Weak Gravity Conjecture may extend to the case of an $\bR$ gauge theory.
    \item We showed that the decompactification limit forces the $d$-dimensional $G_N^{(d)}$ to vanish, but that this still preserves a $d$-dimensional $\bR$ gauge symmetry with no corresponding global symmetries. From the $D$-dimensional perspective, this gauge symmetry is a diffeomorphism in the extra dimension.
\end{itemize}

We wish to discuss some potential future directions and possible extensions of this work:

\begin{itemize}
    \item In this work we considered adding charged scalar fields to break the global symmetries of $\bR$ gauge theory. However, charged fermions would have worked equally well at screening the Wilson lines. Of course, there would be difficulties in producing the cubic interaction terms required to constrain the allowed global charges under the 0-form symmetry, but a full investigation of the fermionic case would be useful.
    \item We mentioned in §\ref{sec:breaking-symmetries} that there was additional evidence for having no non-compact gauge symmetries in quantum gravity, namely the violation of the covariant entropy bound (CEB). We gave a brief argument for how our methods avoid a violation of this bound by decompactifying to a higher-dimensional theory that does \textit{not} violate the bound. Investigating the CEB in this decompactification limit, and its relationship to the Swampland Distance Conjecture \cite{Ooguri:2006in,Ooguri:2018wrx,Grimm:2018ohb,Font:2019cxq} as discussed in \cite{Calderon-Infante:2023ler}, would be interesting.
    \item This work was purely motivated by swampland constrains, to see how to break the global symmetries of non-compact gauge theory by adding uncountably infinitely many particles. This is, of course, not very well motivated phenomenologically. Trying to find a physically relevant scenario in which having a continuum of fields makes sense would be very interesting.
    \item We have only considered the simplest continuous non-compact gauge group in this work, $G=\bR$. An obvious question to ask is whether this works for all non-compact gauge groups. A key method of our work was describing the possible charges of the gauge group, $q\in \widehat{\bR} = \bR$, as momenta in $D$-dimensions. Fourier transforming to position space, we obtain an internal geometry $X_{D-d}= \bR$ whose diffeomorphism group is the gauge group of the $d$-dimensional theory, $\bR$. This is the relationship for general Kaluza-Klein reductions on compact $X_{D-d}$ too. So, if we take general non-compact $G$ gauge theory in $d$-dimensions, there are generically infinite-dimensional unitary representations, and the charges might form a continuum as they did for $G=\bR$. Then, we would need to pick a Haar measure $d\mu(q)$ on $\widehat{G}$, the unitary dual of $G$ that contains these infinite-dimensional representations. This Haar measure for $\widehat{\bR}=\bR$ was just $dq$. Given such a Haar measure, Plancherel's theorem for locally compact groups says there is a unique measure on $G$, $d\nu(g)$, that allows us to perform a Fourier transform from integrals over $\widehat{G}$ with measure $d\mu(q)$ to integrals over $G$ with Fourier transform $d\nu(g)$. For $G=\bR$, this dual measure was just $d\nu(y)=dy$ for $y\in \bR$. Using these ideas to investigate other locally-compact, non-compact groups would make for interesting future work.
\end{itemize}

\section*{Acknowledgments}
We would like to thank Tom Rudelius for helpful discussions. We thank our supervisors Iñaki García Etxebarria and Tom Rudelius for comments on a draft. We are funded by the STFC grant ST/Y509334/1.

\appendix
\section{Other methods of integration to evaluate the Noether current}\label{appendix:other-methods-of-integration}
    In \S\ref{sec:noether-current-integrability}, we argued that the only $Q_q$ obeying \eqref{eq:restrictions-on-Q} that resulted in an integral for the Noether current, and hence the corresponding topological operator, that could be evaluated were those of the form $Q_q=kq$ for some $k\in\bR$.
    This argument was done by assuming Riemann integration.
    The reader may be wondering whether using Riemann integration is too restrictive -- whether using some other form of integration, e.g. Lebesgue integration, would result in a more general class of $Q_q$ having evaluatable Noether currents and consequently a global symmetry.
    In this appendix we outline arguments for why we expect this not to be the case, or rather, why even if it could be evaluated, we expect the resulting topological operator to be the same as for if we had $Q_q=kq$ for some $k\in\bR$, and so not correspond to a global symmetry.

    We first observe that any $Q_q$ obeying \eqref{eq:restrictions-on-Q} can have its domain $\bR$ split up into a collection of sets $D_k$ indexed by $k\in K\subseteq\bR$, with each $D_k$ dense in $\bR$, and $Q_q=kq$ when its domain is restricted to $D_k$.

    The Riemann integrable case is when $\brav{K}=1$ -- in that case, $K=\brac{k}$ and $D_k=\bR$ for some $k\in \bR$.

    If $\brav{K}\neq1$, then the function is in a sense ``trying'' to be all of the functions $f_k\brap{q}=kq$, for all $k \in K$, but which such linear function it is trying to be at a given point $q$ depends upon exactly which dense subset $D_k$ the point $q$ is in.
    Indeed, one way of viewing why it is not Riemann integrable is because the Riemann integral ``wants'' to converge to the integral of each of these linear functions, but can't do so because the resulting limit is different for different linear functions.

    So if we were to find a method of integration that could evaluate the integral when $\brav{K}\neq1$, we'd expect it to be have effectively found a way to ``smear'' over all the linear functions, so the value of the integral is a ``smearing'' over the values of the integrals of the linear functions.
    But since the integral of the linear function $kq$ is $k$ times the integral of the linear function $q$, smearing over the integral of linear functions just gives the integral of some linear function.
    So we would find that the integral of $Q_q$ would by the integral of $k'q$ for some $k'\in\bR$ that comes from the result of this smearing.
    So in fact the resulting topological operator would be that of $Q_q=k'q$ for some $k'\in\bR$, which as explained in §\ref{sec:noether-current-integrability}, does not correspond to a global symmetry.

    To illustrate what we mean by ``smearing'', consider the function $f\brap{x}$, with domain $\bras{0,1}$, and given by:
    \begin{equation}
        f\brap{x} =
        \begin{cases}
            ax &\text{if } x \in A\\
            bx &\text{if } x \in B\,,
        \end{cases}
    \end{equation}
    where $A\subset\bras{0,1}$ and $B=\bras{0,1}\setminus A$ are both dense in $\bras{0,1}$, and $a$, $b$ are positive constants.
    If $a \neq b$, then $f\brap{x}$ is not Riemann integrable over $\bras{0,1}$, for similar reasons to why the general $Q_q$ is not Riemann integrable -- essentially because the integral is trying to converge to the integral of $ax$ and of $bx$, but can't do both as those limits are not the same.

    However, if we let $S=\brac{\brap{r,s} \cap A: r<s\in \bras{0,1}}\cup \brac{\brap{r,s} \cap B: r<s\in \bras{0,1}}$, and perform Lebesgue integration%
    \footnote{
        For a very readable introduction to Lebesgue integration, see \cite{Hogan-Murphy:lebesgue-integration}.
    } %
    with the $\sigma$-algebra given by the smallest $\sigma$-algebra containing $S$, and the measure $\mu$ given by a modified Lebesgue measure:
    \begin{align}
        \mu\brap{\brap{r,s}\cap A} &= \alpha\bras{s-r} \,,\\
        \mu\brap{\brap{r,s}\cap B} &= \beta\bras{s-r}
        \,,
    \end{align}
    where $\alpha\in\bras{0,1}$ and $\alpha+\beta=1$, then we find that evaluating the Lebesgue integral of $f$ over $\bras{u,v}$ is
    \begin{equation}
        \frac12\brap{a\alpha+b\beta}\bras{v^2-u^2} = \frac12 k \bras{v^2-u^2}\,,
    \end{equation}
    where $k=\brap{a\alpha+b\beta}$.
    This is the same as integrating $kx$ (using either Lebesgue integration with $\mu$ or Riemann integration) over the same interval.

    What has this integration done? It has ``smeared'' over the two regions where $f$ ``tries to be'' the two linear functions $ax$ and $bx$ by giving weights $\alpha$ and $\beta$ to the two regions. The consequence is that the integral of $f$ becomes that of the linear function $kx$, where $k$ is found by ``smearing'' $a$ and $b$ by weighting them by the weights $\alpha$ and $\beta$ of the corresponding intervals.

\section{Propagators and the species scale}\label{appendix:extra-dim-propagators-and-species-scale}
    We began §\ref{sec:extra-dimensions} by discussing the species scale, and how an EFT with a countable infinity of fields has an arbitrarily small UV cut-off scale. However, the EFT we are considering in this work has a \textit{continuum} of fields, and so it may be naïve to assume that we can simply take $N \rightarrow \infty$. Indeed, we will show in this appendix that having a \textit{continuum} of fields means one cannot derive a $d$-dimensional species scale. However, we will show that treating the EFT as a $D$-dimensional theory gives a valid perturbative EFT with a well-defined UV cut-off. We will work in $d=4$ for simplicity in this section.

    Consider an action
    \begin{equation}
        S = \int d^4x \sqrt{-g} \left(\frac{1}{G_N}\cR + g^{\mu \nu} \partial_\mu \phi^\dagger \partial_\nu \phi + m^2 \phi^\dagger \phi \right)\,,
    \end{equation}
    where $G_N = \frac{1}{M_\mathrm{Pl;4}}$, the $4$-dimensional Newton's constant. One defines the graviton as a perturbation around the flat space inverse metric,
    \begin{equation}
        g^{\mu\nu} \approx \eta^{\mu\nu} + h^{\mu\nu}
    \end{equation}
    such that
    \begin{equation}
        \sqrt{-g} \approx 1 + \eta_{\mu\nu} h^{\mu\nu} + \dots
    \end{equation}
    Then, from e.g. \cite{Jimu:2024xqm} we can obtain the interaction vertices
    \begin{align}
        v^{\mu\nu} &= \begin{tikzpicture}[baseline=(v.base)]
    \begin{feynman}
        \vertex (v);
        \vertex [above left=1.5cm and 1.5cm of v] (s1) [particle=\(\phi\)];
        \vertex [below left=1.5cm and 1.5cm of v] (s2) [particle=\(\phi^\dagger\)];
        \vertex [right=2cm of v] (g) [particle=g];
        \diagram*{
            (s1) -- [solid, momentum=\(p_1\)] (v) -- [solid, momentum=\(p_2\)] (s2),
            (v) -- [gluon, momentum'=\(k\), edge label=\(\mu\nu\)] (g),
        };
    \end{feynman}
    \end{tikzpicture}  = -\frac{i}{M_\mathrm{Pl;4}} \{ p_1^{(\mu}p_2^{\nu)} - \eta ^{\mu\nu} (p_1 \cdot p_2 + m^2)    \}\,, \\ \nonumber \\
        w^{\mu\nu,\rho\sigma} &= \begin{tikzpicture}[baseline=(v.base), scale=0.8]
    \begin{feynman}
        \vertex (v) [dot];
        \vertex [above left=1.5cm and 1.5cm of v] (s1) [particle=\(\phi\)];
        \vertex [below left=1.5cm and 1.5cm of v] (s2) [particle=\(\phi^\dagger\)];
        \vertex [above right=1.5cm and 2cm of v] (g1) [particle=g^a];
        \vertex [below right=1.5cm and 2cm of v] (g2) [particle=g^b];
        \diagram*{
            (s1) -- [solid, momentum=\(p\)] (v) -- [solid, momentum=\(p\)] (s2),
            (v) -- [gluon, momentum=\(k\), edge label'=\(\mu\nu\)] (g1),
            (v) -- [gluon, reversed momentum'=\(k\), edge label=\(\rho\sigma\)] (g2),
        };
    \end{feynman}
    \end{tikzpicture} = \frac{i}{2M_\mathrm{Pl;4}} \{ p^\mu p^\nu \eta^{\rho\sigma} + p^\rho p^\sigma \eta^{\mu\nu} - (\frac{1}{2} \eta^{\mu\nu} \eta^{\rho\sigma} + \eta^{\mu(\rho}\eta^{\sigma)\nu}) (p^2 +m^2)) \} \,.
    \end{align}
    The tree-level graviton propagator is
    \begin{equation}
        G_{(0)}^{\mu\nu,\rho\sigma}(p) = \frac{i}{p^2} \{ \eta^{\mu\rho}\eta^{\nu\sigma} + \eta^{\nu\rho}\eta^{\mu\sigma} - \eta^{\mu\nu}\eta^{\rho\sigma} \} \equiv \frac{i H^{\mu\nu,\rho\sigma}}{p^2}\,.
    \end{equation}
    We can then compute the 1-loop contribution to the graviton propagator as:
    \begin{equation}
        G_{\mu\nu,\rho\sigma}(p) \approx \feynmandiagram [small] {
        a -- [gluon, momentum=\(p\)] b;
    }; + \begin{tikzpicture}[baseline=(a.base), scale=0.8]
    \begin{feynman}
    \vertex (a);
    \vertex [right=1cm of a] (b);
    \vertex [right=1cm of b] (c);
    \vertex [right=1cm of c] (d);
    \diagram[small]{
    (a) -- [gluon, momentum=\(p\)] (b),
    (b) -- [solid, half left, looseness=1.5, momentum={[arrow distance = 1mm, arrow shorten =0.1mm]\(k\)}] (c),
    (c) -- [solid, half left, looseness=1.5, momentum={[arrow distance = 1mm, arrow shorten =0.1mm]\(k-p\)}] (b),
    (c) -- [gluon, momentum=\(p\)] (d),
    };
    \end{feynman}
    \end{tikzpicture} + \begin{tikzpicture}[baseline=(a.base), scale=0.8]
    \begin{feynman}
    \vertex (a);
    \vertex [right=1cm of a] (b);
    \vertex [right=1cm of b] (c);
    \vertex [above=1cm of b] (d);
    \diagram[small]{
    (a)   -- [gluon, momentum'=\(p\)] (b) ,
    (b) -- [solid, half left, looseness=0.75, momentum={[arrow distance = 1mm, arrow shorten =0.1mm]\(k\)}] (d),
    (d) -- [solid, half left, looseness=0.75] (b),
    (b) -- [gluon, momentum'=\(p\)] (c),
    };
    \end{feynman}
    \end{tikzpicture} \,,
    \end{equation}
    where we have omitted Lorentz indices in the diagrams.

    Of course, the case of interest to us is when the scalars are \textit{charged}. The action will then imply we have the photon-graviton-scalar vertices
    \begin{equation}
        \begin{tikzpicture}[baseline=(v.base), scale=0.8]
    \begin{feynman}
        \vertex (v) [dot];
        \vertex [above left=1.5cm and 1.5cm of v] (s1) [particle=\(\phi\)];
        \vertex [below left=1.5cm and 1.5cm of v] (s2) [particle=\(\phi^\dagger\)];
        \vertex [above right=1.5cm and 2cm of v] (g1) [particle=g^a];
        \vertex [below right=1.5cm and 2cm of v] (g2) [particle=g^b];
        \vertex [left=3cm of v] (phL);
        \vertex [right=3cm of v] (phR);
        \diagram*{
            (s1) -- [solid] (v) -- [solid] (s2),
            (v) -- [gluon] (g1),
            (v) -- [gluon] (g2),
            (phL) -- [photon] (v),
        };
    \end{feynman}
    \end{tikzpicture} \text{ and \ } \begin{tikzpicture}[baseline=(v.base), scale=0.8]
    \begin{feynman}
        \vertex (v) [dot];
        \vertex [above left=1.5cm and 1.5cm of v] (s1) [particle=\(\phi\)];
        \vertex [below left=1.5cm and 1.5cm of v] (s2) [particle=\(\phi^\dagger\)];
        \vertex [above right=1.5cm and 2cm of v] (g1) [particle=g^a];
        \vertex [below right=1.5cm and 2cm of v] (g2) [particle=g^b];
        \vertex [left=3cm of v] (phL);
        \vertex [right=3cm of v] (phR);
        \diagram*{
            (s1) -- [solid] (v) -- [solid] (s2),
            (v) -- [gluon] (g1),
            (v) -- [gluon] (g2),
            (phL) -- [photon] (v) -- [photon] (phR),
        };
    \end{feynman}
    \end{tikzpicture}\,,
    \end{equation}
    but these will not contribute to the graviton self-energy at 1-loop order -- therefore we can ignore these contributions.

    The above loop diagrams contribute to the graviton self-energy as
    \begin{align}
        \begin{tikzpicture}[baseline=(a.base), scale=0.8]
    \begin{feynman}
    \vertex (a);
    \vertex [right=1cm of a] (b);
    \vertex [right=1cm of b] (c);
    \vertex [right=1cm of c] (d);
    \diagram[small]{
    (a) -- [gluon, momentum=\(p\), edge label'=\(\mu\nu\)] (b),
    (b) -- [solid, half left, looseness=1.5, momentum={[arrow distance = 1mm, arrow shorten =0.1mm]\(k\)}] (c),
    (c) -- [solid, half left, looseness=1.5, momentum={[arrow distance = 1mm, arrow shorten =0.1mm]\(k-p\)}] (b),
    (c) -- [gluon, momentum=\(p\), edge label'=\(\rho\sigma\)] (d),
    };
    \end{feynman}
    \end{tikzpicture} & = \int \dbar k \cdot v_{\mu\nu}(k,p) v_{\rho\sigma}(k,p) \cdot \frac{i}{k^2-m_q^2} \cdot \frac{i}{(k-p)^2-m_q^2} \,,\\
    \begin{tikzpicture}[baseline=(a.base), scale=0.8]
    \begin{feynman}
    \vertex (a);
    \vertex [right=1cm of a] (b);
    \vertex [right=1cm of b] (c);
    \vertex [above=1cm of b] (d);
    \diagram[small]{
    (a)   -- [gluon, edge label'=\(\mu\nu\)] (b) ,
    (b) -- [solid, half left, looseness=0.75, momentum={[arrow distance = 1mm, arrow shorten =0.1mm]\(k\)}] (d),
    (d) -- [solid, half left, looseness=0.75] (b),
    (b) -- [gluon, momentum={[arrow distance = 2mm, arrow shorten =0.1mm]\(p\)}, edge label'=\(\rho\sigma\)] (c),
    };
    \end{feynman}
    \end{tikzpicture} &=  \int \dbar k \cdot  w_{\mu\nu,\rho\sigma}(k) \cdot \frac{i}{k^2 -m_q^2}\,.
    \end{align}
    Each $\phi_q$ will contribute to the self-energy in this way, so that the total graviton self-energy is
    \begin{equation}
        \Pi_{\mu\nu,\rho\sigma}(p) = \int  dq \cdot \Pi^{(q)}_{\mu\nu,\rho\sigma}(p) \,,
    \end{equation}
    where $\Pi^{(q)}_{\mu\nu,\rho\sigma}(p)$ is the sum of the above loop diagrams.

    With a countable number of scalars we would not have an integral here, but a discrete sum:
    \begin{equation}
        \Pi_{\mu\nu,\rho\sigma}(p) =  \sum_{n=1}^N \Pi^{(n)}_{\mu\nu,\rho\sigma}(p) \,.
    \end{equation}
    Often one can approximate $m_n \approx m_{n'}$ so that the self-energy is simply \cite{Castellano:2022bvr}
    \begin{equation}
         \Pi_{\mu\nu,\rho\sigma}(p) \approx \frac{N p^4}{M_\mathrm{Pl;4}^2} \log{\frac{-p^2}{\mu^2}} \cdot H_{\mu\nu,\rho\sigma}
    \end{equation}
    after computing the loop integrals above. Then, one considers the resummation of the graviton propagator
    \begin{equation}
        G_{\mu\nu,\rho\sigma}(p) = \frac{iH_{\mu\nu,\rho\sigma}}{p^2 \cdot(1-\frac{Np^2}{M_\mathrm{Pl;4}^2}\log{\frac{-p^2}{\mu^2}})} \,.
    \end{equation}
    The species scale is then defined to be the energy scale where perturbation theory breaks down, which is when the denominator approaches zero:
    \begin{equation}
        \Lambda_\mathrm{sp} \approx \frac{M_\mathrm{Pl;4}}{\sqrt{N}} \,.
    \end{equation}
    Therefore, the number of light species, i.e. those that are not heavy enough to integrate out, lowers the UV cut-off of the EFT.

    To return to our case of a continuum of fields, the self-energy of the graviton will involve integrals of the form
    \begin{equation}
        \int dq \cdot \int d^4k \frac{i}{k^2+m_q^2}\,.
    \end{equation}
    We may choose some 4d renormalisation scheme to render the integral over $k$ finite, to obtain the 1-loop contribution to the self-energy coming from a single field. However, considering every contribution from the continuum of fields by integrating over $q$ will make this 1-loop contribution diverge. Therefore, we cannot do perturbation theory at \textit{any} energy scale, so the UV cut-off is always zero.

    However, if we consider the the integral above after defining the 5-momentum $P_M = (k_\mu, m_q)$, then we obtain a \textit{5d propagator of a massless field}:
    \begin{equation}
        \int dq \cdot \int d^4k \frac{i}{k^2+m_q^2} =\int d^5 P \cdot \frac{i}{P^2} \,.
    \end{equation}
    That is, the 1-loop contributions to the 4d graviton from a continuum of fields with mass spectra $m_q^2$ can be viewed instead as the contribution from the interaction of the single 5d scalar considered in \eqref{eq:Einstein-Scalar}. We can then use 5d renormalisation schemes to render this contribution finite in the 5d EFT and define a 5d species scale as above, in this case with $N=1$ due to there being only a single 5d scalar.

\section{Matching mass dimensions between the \texorpdfstring{$D$}{D} and \texorpdfstring{$d$}{d} dimensional theories}\label{appendix:matching-mass-dimensions}
    In this appendix, we demonstrate how to rescale the theory from \S\ref{sec:extra-dimensions} so that we arrive at \eqref{eq:scalar-decompactification} without an extra power of $\frac1{\sqrt\lambda}$ that would be there in the original normalisation of quantities, and how to translate between $d$-dimensional and $D$ dimensional quantities while ensuring they have the correct mass dimension in each theory.
    These issues are not unique to Kaluza-Klein reductions on $\bR$ -- at least some variant of them occurs for Kaluza-Klein reductions on $S^1$ -- but in this appendix we make their remedies explicit.

    For clarity, we denote $d$-dimensional quantities with a $^\brap{d}$ and $D$-dimensional quantities with a $^\brap{D}$.

    From \eqref{eq:D-dim-metric}, we have
    \begin{equation*}
        {G^{\brap{D}}}_{MN}\brap{X} =
		\begin{pmatrix}
			g^{\brap{D}}_{\mu\nu}\brap{x} + \frac1{\lambda^{\brap{D}}}A^{\brap{D}}_\mu\brap{x}A^{\brap{D}}_\nu\brap{x} & \frac1{\lambda^{\brap{D}}}A^{\brap{D}}_\mu\brap{x}\\
			\frac1{\lambda^{\brap{D}}}A^{\brap{D}}_\nu\brap{x} & \frac1{\lambda^{\brap{D}}}
		\end{pmatrix}
        \,.
    \end{equation*}
    Metric components are mass dimensionless, so we find
    \begin{equation}
        \bras{g^\brap{D}_{\mu\nu}}
        =
        \bras{\lambda^\brap{D}}
        =
        \bras{A^\brap{D}_\mu}
        =
        0
        \,,
    \end{equation}
    whereas $q^{\brap{D}}$ is a momentum, so we find
    \begin{equation}
        \bras{q^{\brap{D}}}=1
        \,,
    \end{equation}
    and as $\Phi^{\brap{D}}$ is a scalar field in $D$-dimensions, it has mass dimension
    \begin{equation}
        \bras{\Phi^{\brap{D}}}=\frac{D-2}{2}
        \,.
    \end{equation}

    On the other hand, in $d$ dimensions, $q$ is a charge, so we require
    \begin{equation}
        \bras{q^{\brap{d}}}=0
        \,.
    \end{equation}
    We therefore define a mass scale $\mu$ by
    \begin{equation}\label{eq:defining-mass-scale-mu}
        q^{\brap{D}}=\mu \cdot q^{\brap{d}}
        \,.
    \end{equation}
    Then from $A$ multiplying $q$ in $D\phi$, we find
    \begin{equation}
        A^{\brap{D}}_\nu=\frac1\mu \cdot A^{\brap{d}}_\nu\,,
    \end{equation}
    which then satisfies
    \begin{equation}
        \bras{A^\brap{d}_\mu}=1\,,
    \end{equation}
    as is required of a gauge field.
    Also, from $m^2=M^2+\lambda q^2$, we find
    \begin{equation}
        \lambda^{\brap{D}}=\frac1{\mu^2} \cdot \lambda^{\brap{d}}
        \,,
    \end{equation}
    yielding
    \begin{equation}
        \bras{\lambda^\brap{d}_\mu}=2\,.
    \end{equation}

    $g^\brap{D}_{\alpha\beta}$ already has the correct mass dimensions, and so we take
    \begin{equation}
        g^\brap{d}_{\alpha\beta}=g^\brap{D}_{\alpha\beta}\,.
    \end{equation}

    Lastly, the redefinition of $\lambda$ and $\dbar q$ in \eqref{eq:scalar-decompactification} leaves left-over factors of $\mu$, and we have an overall factor of $\frac1{\sqrt{\lambda}}$ in going from $\sqrt{-G}$ to $\sqrt{-g}$ when deriving \eqref{eq:scalar-decompactification}.
    Defining
    \begin{equation}
        \phi^{\brap{D}}=\frac{\brap{\lambda^{\brap{d}}}^{\frac14}}{\mu} \cdot \phi^{\brap{d}}\,,
    \end{equation}
    we arrive at
    \begin{equation}
        \bras{\phi^{\brap{d}}}=\frac{d-2}{2}
        \,,
    \end{equation}
    as required of a scalar in $d$-dimensions. Here we are treating $\lambda$ ($d$-dimensional or $D$-dimensional) as independent of position. If it had spatial dependence -- i.e. if it were a field -- this redefinition of $\phi$ would induce additional terms in the Lagrangian involving $\lambda$.

    Note that these re-scalings have no effect on the $\cR$ and $\frac1{2\lambda}F_{\mu\nu}F^{\mu\nu}$ terms in \eqref{eq:Einstein-Maxwell-Dilaton}, as the former depends only on quantities that aren't re-scaled, and the re-scalings of $\lambda$ and of $A$ cancel each other in the latter.

    So, starting with the $D$-dimensional quantities with correct mass dimensions, if we define the $d$-dimensional quantities as above, we arrive at \eqref{eq:scalar-decompactification}, with no extra factor of $\frac1{\sqrt\lambda}$, and with all $d$-dimensional quantities having the correct mass dimension.

    But what is the mass scale $\mu$?

    For U(1) gauge theory, \eqref{eq:defining-mass-scale-mu} becomes
    \begin{equation}
        q^{\brap{d}}=\frac1\mu q^{\brap{D}} = \frac{n}{\mu R}\,,
    \end{equation}
    where $n$ could be any integer, and $R$ is the radius of the compactification circle.
    Imposing the standard normalisation that the smallest non-zero charge is 1 enforces that $\mu=\frac1R$ -- i.e. $\mu$ is the mass scale associated to the circle radius.

    But for $\bR$ gauge theory there is no minimum non-zero charge, and so we can't use this to fix $\mu$.

    We however claim that any finite positive $\mu$ is physically equivalent, as it only serves to correct mass dimension from $D$-dimensions to $d$ dimensions -- for fields it re-scales them as above, and for coupling constants it corrects their mass dimension.
    Indeed, any Lagrangian theory is invariant under re-scaling its momenta by a positive constant if you also allow all coupling constants to re-scale, and that is effectively what we are doing here.

\section{Breaking the symmetry with charges in a finite interval \texorpdfstring{$\brap{-L,L}$}{(-L,L)}}\label{appendix:finite-charge-interval}
    In \S\ref{sec:constraining-Q-by-interactions} and elsewhere in the main body of the text, we added a fundamental field $\phi_q$ for every charge $q\in\bR$.
    Every Wilson line $\cW_q$ was then screened by the field $\phi_q$, breaking the 1-form symmetry.
    Further, having uncountably infinitely many charges converted the usual sum over charges to an integral over charges, allowing us to eliminate the candidate global 0-form symmetries via integrability.

    But to what extent was \emph{all} of $\bR$ necessary -- could we perhaps just consider a subset of $\bR$ and achieve the same task?

    For the integrability argument, we needed enough fields to be forced to make sense of them via an integral over charges instead of a sum. Restricting $q$ to be in some dense subset of the reals, e.g. $\bQ$, would also enforce this, but in general would leave some of the Wilson lines unscreened -- in the case of $\bQ$, all of the Wilson lines of irrational charge.

    One possibility is to take the $\bR$ and remove discrete points -- as long as we don't remove too many, we can still screen all the Wilson lines by taking products of fields. We have not investigated this possibility much, as it could get quite messy. We expect however that at least in some cases this can be reduced to the previous case by ``filling-in'' the missing $\phi_q$'s based on continuity arguments for $\phi_q$.

    Another possibility, which we shall consider further, is to instead consider charges in some finite interval, say
    \begin{equation}
        \cI := \brap{-L,L}
        \,,
    \end{equation}
    for some positive constant $L$.
    Repeating the arguments of \S\ref{sec:constraining-Q-by-interactions}-\S\ref{sec:noether-current-integrability} whilst being careful to never assume the existence of a charge outside of $\cI$, we find that the integrability arguments follow through --
    we do not introduce a 0-form global symmetry.

    Further, for any $q\in\bR$, there is some finite integer $n$ such that $\frac{q}{n}\in\cI$, and so the Wilson line is screened by the operator $\phi_{q/n}^n$.
    Hence all the Wilson lines are screened, and so there is no associated global 1-form symmetry.

    So requiring charges in $\cI$ is sufficient for breaking the global symmetry.

    We demonstrated in \S\ref{sec:extra-dimensions} how the theory with fields for all $q\in\bR$, while hard to make sense of in $d$ dimensions due to its infinitely many fields, can be seen as a theory in one dimension higher with finitely many fields.
    The theory restricted to $q\in\cI$ also has infinitely many fields in $d$ dimensions, so how do we make sense of it?

    In this case, the same arguments do \emph{not} quite work -- recasting in terms of a theory in one dimension higher relied on the integral over $q$ being over all $\bR$ to invoke an inverse Fourier transform; this breaks down if we restrict the integral over $q$ to be in $\cI$.
    We can however modify the arguments slightly to try to make sense of the theory.

    If we consider the logic of §\ref{sec:KK-reduction} in reverse, such that we never introduce a $y$ coordinate, we can view the $d$-dimensional scalar kinetic term in \textit{momentum space} as
    \begin{eqnarray}
        \int_{\cM^d} d^dx \cdot \dbar^dp_1 \cdot \dbar^dp_2
        \sqrt{-G\brap{x}}
        \int_{-L}^L\dbar q_1 \cdot \int_{-L}^L \dbar q_2 \cdot
        \brap{2\pi}\delta\brap{q_1-q_2}
        \cdot e^{i\brap{p_1-p_2}\cdot x}
        \nonumber\\ \times
        \tilde\phi^\dagger_{q_1}\brap{p_1}  \cdot \tilde\phi_{q_2}\brap{p_2}
        \cdot
        \bras{
            \brap{P_1}_MG^{MN}\brap{x}\brap{P_2}_N
            + M^2
        }
        \,,
    \end{eqnarray}
    where we define the $D$-momenta $(P_i)_M=((p_i)_\mu,q_i)$ and use the $D$-dimensional metric $G_{MN}$ as in \eqref{eq:D-dim-metric}. We could also Fourier transform the metric to $d$-dimensional momentum space and then evaluate the integral over $x$ to eliminate $x$ from the integral in favour of momentum-conserving delta-functions, fully converting the expression to momentum space.
    Then, one can view the integral over $(-L,L)$ to be a momentum cut-off in the $M=D-1$ direction. This cut-off means we cannot introduce the $y$ coordinate and use an inverse Fourier transform to write this as a $D$-dimensional action in \textit{position space.} When we send $L\rightarrow \infty$, so that we are considering fields $\phi_q$ for all $q\in \bR$, we can interpret this as removing the momentum cut-off in the $D$-direction, which allows us to write the above in position space.

\bibliographystyle{JHEP}
\baselineskip=.95\baselineskip
\bibliography{refs}

\providecommand{\href}[2]{#2}\begingroup\raggedright\begin{thebibliography}{10}

\bibitem{Vafa:2005ui}
C.~Vafa, \emph{{The String landscape and the swampland}},
  \href{https://arxiv.org/abs/hep-th/0509212}{{\ttfamily hep-th/0509212}}.

\bibitem{Rudelius:2024vmc}
T.~Rudelius, \emph{{A symmetry-centric perspective on the geometry of the
  string landscape and the swampland}},
  \href{https://doi.org/10.1142/S0218271824410037}{\emph{Int. J. Mod. Phys. D}
  {\bfseries 33} (2024) 2441003}
  [\href{https://arxiv.org/abs/2405.12980}{{\ttfamily 2405.12980}}].

\bibitem{vanBeest:2021lhn}
M.~van Beest, J.~Calder{\'o}n-Infante, D.~Mirfendereski and I.~Valenzuela,
  \emph{{Lectures on the Swampland Program in String Compactifications}},
  \href{https://doi.org/10.1016/j.physrep.2022.09.002}{\emph{Phys. Rept.}
  {\bfseries 989} (2022) 1} [\href{https://arxiv.org/abs/2102.01111}{{\ttfamily
  2102.01111}}].

\bibitem{Rudelius:2024mhq}
T.~Rudelius, \emph{{An Introduction to the Weak Gravity Conjecture}},
  \href{https://doi.org/10.1080/00107514.2024.2391206}{\emph{Contemp. Phys.}
  {\bfseries 1} (2024) 14} [\href{https://arxiv.org/abs/2409.02161}{{\ttfamily
  2409.02161}}].

\bibitem{Harlow:2022ich}
D.~Harlow, B.~Heidenreich, M.~Reece and T.~Rudelius, \emph{{Weak gravity
  conjecture}}, \href{https://doi.org/10.1103/RevModPhys.95.035003}{\emph{Rev.
  Mod. Phys.} {\bfseries 95} (2023) 035003}
  [\href{https://arxiv.org/abs/2201.08380}{{\ttfamily 2201.08380}}].

\bibitem{Palti:2019pca}
E.~Palti, \emph{{The Swampland: Introduction and Review}},
  \href{https://doi.org/10.1002/prop.201900037}{\emph{Fortsch. Phys.}
  {\bfseries 67} (2019) 1900037}
  [\href{https://arxiv.org/abs/1903.06239}{{\ttfamily 1903.06239}}].

\bibitem{Brennan:2017rbf}
T.D.~Brennan, F.~Carta and C.~Vafa, \emph{{The String Landscape, the Swampland,
  and the Missing Corner}},
  \href{https://doi.org/10.22323/1.305.0015}{\emph{PoS} {\bfseries TASI2017}
  (2017) 015} [\href{https://arxiv.org/abs/1711.00864}{{\ttfamily
  1711.00864}}].

\bibitem{Grana:2021zvf}
M.~Gra{\~n}a and A.~Herr{\'a}ez, \emph{{The Swampland Conjectures: A Bridge
  from Quantum Gravity to Particle Physics}},
  \href{https://doi.org/10.3390/universe7080273}{\emph{Universe} {\bfseries 7}
  (2021) 273} [\href{https://arxiv.org/abs/2107.00087}{{\ttfamily
  2107.00087}}].

\bibitem{Reece:2023czb}
M.~Reece, \emph{{TASI Lectures: (No) Global Symmetries to Axion Physics}},
  \href{https://doi.org/10.22323/1.439.0008}{\emph{PoS} {\bfseries TASI2022}
  (2024) 008} [\href{https://arxiv.org/abs/2304.08512}{{\ttfamily
  2304.08512}}].

\bibitem{Hawking:1975vcx}
S.W.~Hawking, \emph{{Particle Creation by Black Holes}},
  \href{https://doi.org/10.1007/BF02345020}{\emph{Commun. Math. Phys.}
  {\bfseries 43} (1975) 199}.

\bibitem{Zeldovich:1976vw}
Y.B.~Zeldovich, \emph{{Charge Asymmetry of the Universe Due to Black Hole
  Evaporation and Weak Interaction Asymmetry}}, {\emph{Pisma Zh. Eksp. Teor.
  Fiz.} {\bfseries 24} (1976) 29}.

\bibitem{Zeldovich:1977mk}
Y.B.~Zeldovich, \emph{{The Gravitation, Charges, Cosmology and Coherence}},
  \href{https://doi.org/10.3367/UFNr.0123.197711d.0487}{\emph{Usp. Fiz. Nauk}
  {\bfseries 123} (1977) 487}.

\bibitem{Banks:1988yz}
T.~Banks and L.J.~Dixon, \emph{{Constraints on String Vacua with Space-Time
  Supersymmetry}},
  \href{https://doi.org/10.1016/0550-3213(88)90523-8}{\emph{Nucl. Phys. B}
  {\bfseries 307} (1988) 93}.

\bibitem{Giddings:1988cx}
S.B.~Giddings and A.~Strominger, \emph{{Loss of incoherence and determination
  of coupling constants in quantum gravity}},
  \href{https://doi.org/10.1016/0550-3213(88)90109-5}{\emph{Nucl. Phys. B}
  {\bfseries 307} (1988) 854}.

\bibitem{Abbott:1989jw}
L.F.~Abbott and M.B.~Wise, \emph{{Wormholes and Global Symmetries}},
  \href{https://doi.org/10.1016/0550-3213(89)90503-8}{\emph{Nucl. Phys. B}
  {\bfseries 325} (1989) 687}.

\bibitem{Coleman:1989zu}
S.R.~Coleman and K.-M.~Lee, \emph{{WORMHOLES MADE WITHOUT MASSLESS MATTER
  FIELDS}}, \href{https://doi.org/10.1016/0550-3213(90)90149-8}{\emph{Nucl.
  Phys. B} {\bfseries 329} (1990) 387}.

\bibitem{Kallosh:1995hi}
R.~Kallosh, A.D.~Linde, D.A.~Linde and L.~Susskind, \emph{{Gravity and global
  symmetries}}, \href{https://doi.org/10.1103/PhysRevD.52.912}{\emph{Phys. Rev.
  D} {\bfseries 52} (1995) 912}
  [\href{https://arxiv.org/abs/hep-th/9502069}{{\ttfamily hep-th/9502069}}].

\bibitem{Banks:2010zn}
T.~Banks and N.~Seiberg, \emph{{Symmetries and Strings in Field Theory and
  Gravity}}, \href{https://doi.org/10.1103/PhysRevD.83.084019}{\emph{Phys. Rev.
  D} {\bfseries 83} (2011) 084019}
  [\href{https://arxiv.org/abs/1011.5120}{{\ttfamily 1011.5120}}].

\bibitem{Harlow:2018tng}
D.~Harlow and H.~Ooguri, \emph{{Symmetries in quantum field theory and quantum
  gravity}}, \href{https://doi.org/10.1007/s00220-021-04040-y}{\emph{Commun.
  Math. Phys.} {\bfseries 383} (2021) 1669}
  [\href{https://arxiv.org/abs/1810.05338}{{\ttfamily 1810.05338}}].

\bibitem{Harlow:2020bee}
D.~Harlow and E.~Shaghoulian, \emph{{Global symmetry, Euclidean gravity, and
  the black hole information problem}},
  \href{https://doi.org/10.1007/JHEP04(2021)175}{\emph{JHEP} {\bfseries 04}
  (2021) 175} [\href{https://arxiv.org/abs/2010.10539}{{\ttfamily
  2010.10539}}].

\bibitem{Chen:2020ojn}
Y.~Chen and H.W.~Lin, \emph{{Signatures of global symmetry violation in
  relative entropies and replica wormholes}},
  \href{https://doi.org/10.1007/JHEP03(2021)040}{\emph{JHEP} {\bfseries 03}
  (2021) 040} [\href{https://arxiv.org/abs/2011.06005}{{\ttfamily
  2011.06005}}].

\bibitem{Hsin:2020mfa}
P.-S.~Hsin, L.V.~Iliesiu and Z.~Yang, \emph{{A violation of global symmetries
  from replica wormholes and the fate of black hole remnants}},
  \href{https://doi.org/10.1088/1361-6382/ac2134}{\emph{Class. Quant. Grav.}
  {\bfseries 38} (2021) 194004}
  [\href{https://arxiv.org/abs/2011.09444}{{\ttfamily 2011.09444}}].

\bibitem{Yonekura:2020ino}
K.~Yonekura, \emph{{Topological violation of global symmetries in quantum
  gravity}}, \href{https://doi.org/10.1007/JHEP09(2021)036}{\emph{JHEP}
  {\bfseries 09} (2021) 036}
  [\href{https://arxiv.org/abs/2011.11868}{{\ttfamily 2011.11868}}].

\bibitem{McNamara:2021cuo}
J.~McNamara, \emph{{Gravitational Solitons and Completeness}},
  \href{https://arxiv.org/abs/2108.02228}{{\ttfamily 2108.02228}}.

\bibitem{Gaiotto:2014kfa}
D.~Gaiotto, A.~Kapustin, N.~Seiberg and B.~Willett, \emph{{Generalized Global
  Symmetries}}, \href{https://doi.org/10.1007/JHEP02(2015)172}{\emph{JHEP}
  {\bfseries 02} (2015) 172} [\href{https://arxiv.org/abs/1412.5148}{{\ttfamily
  1412.5148}}].

\bibitem{Rudelius:2020orz}
T.~Rudelius and S.-H.~Shao, \emph{{Topological Operators and Completeness of
  Spectrum in Discrete Gauge Theories}},
  \href{https://doi.org/10.1007/JHEP12(2020)172}{\emph{JHEP} {\bfseries 12}
  (2020) 172} [\href{https://arxiv.org/abs/2006.10052}{{\ttfamily
  2006.10052}}].

\bibitem{Heidenreich:2021xpr}
B.~Heidenreich, J.~McNamara, M.~Montero, M.~Reece, T.~Rudelius and
  I.~Valenzuela, \emph{{Non-invertible global symmetries and completeness of
  the spectrum}}, \href{https://doi.org/10.1007/JHEP09(2021)203}{\emph{JHEP}
  {\bfseries 09} (2021) 203}
  [\href{https://arxiv.org/abs/2104.07036}{{\ttfamily 2104.07036}}].

\bibitem{Hopkins:2002rd}
M.J.~Hopkins and I.M.~Singer, \emph{{Quadratic functions in geometry, topology,
  and M theory}}, {\emph{J. Diff. Geom.} {\bfseries 70} (2005) 329}
  [\href{https://arxiv.org/abs/math/0211216}{{\ttfamily math/0211216}}].

\bibitem{Sulejmanpasic:2019ytl}
T.~Sulejmanpasic and C.~Gattringer, \emph{{Abelian gauge theories on the
  lattice: $\theta$-Terms and compact gauge theory with(out) monopoles}},
  \href{https://doi.org/10.1016/j.nuclphysb.2019.114616}{\emph{Nucl. Phys. B}
  {\bfseries 943} (2019) 114616}
  [\href{https://arxiv.org/abs/1901.02637}{{\ttfamily 1901.02637}}].

\bibitem{Gorantla:2021svj}
P.~Gorantla, H.T.~Lam, N.~Seiberg and S.-H.~Shao, \emph{{A modified Villain
  formulation of fractons and other exotic theories}},
  \href{https://doi.org/10.1063/5.0060808}{\emph{J. Math. Phys.} {\bfseries 62}
  (2021) 102301} [\href{https://arxiv.org/abs/2103.01257}{{\ttfamily
  2103.01257}}].

\bibitem{MathWorld:hamel:basis}
K.~O'Bryant, ``Hamel basis.'' From MathWorld--A Wolfram Resource, created by
  Eric W. Weisstein. \url{https://mathworld.wolfram.com/HamelBasis.html}.
\newblock Accessed: 2026-01-20.

\bibitem{nlab:basis_of_a_vector_space}
{nLab authors}, ``basis of a vector space.''
  \url{https://ncatlab.org/nlab/show/basis+of+a+vector+space}.
\newblock
  \href{https://ncatlab.org/nlab/revision/basis+of+a+vector+space/9}{Revision
  9}, accessed 2026-01-20.

\bibitem{Arkani-Hamed:2005zuc}
N.~Arkani-Hamed, S.~Dimopoulos and S.~Kachru, \emph{{Predictive landscapes and
  new physics at a TeV}},
  \href{https://arxiv.org/abs/hep-th/0501082}{{\ttfamily hep-th/0501082}}.

\bibitem{Distler:2005hi}
J.~Distler and U.~Varadarajan, \emph{{Random polynomials and the friendly
  landscape}},  \href{https://arxiv.org/abs/hep-th/0507090}{{\ttfamily
  hep-th/0507090}}.

\bibitem{Dimopoulos:2005ac}
S.~Dimopoulos, S.~Kachru, J.~McGreevy and J.G.~Wacker, \emph{{N-flation}},
  \href{https://doi.org/10.1088/1475-7516/2008/08/003}{\emph{JCAP} {\bfseries
  08} (2008) 003} [\href{https://arxiv.org/abs/hep-th/0507205}{{\ttfamily
  hep-th/0507205}}].

\bibitem{Dvali:2007hz}
G.~Dvali, \emph{{Black Holes and Large N Species Solution to the Hierarchy
  Problem}}, \href{https://doi.org/10.1002/prop.201000009}{\emph{Fortsch.
  Phys.} {\bfseries 58} (2010) 528}
  [\href{https://arxiv.org/abs/0706.2050}{{\ttfamily 0706.2050}}].

\bibitem{Dvali:2007wp}
G.~Dvali and M.~Redi, \emph{{Black Hole Bound on the Number of Species and
  Quantum Gravity at LHC}},
  \href{https://doi.org/10.1103/PhysRevD.77.045027}{\emph{Phys. Rev. D}
  {\bfseries 77} (2008) 045027}
  [\href{https://arxiv.org/abs/0710.4344}{{\ttfamily 0710.4344}}].

\bibitem{Castellano:2022bvr}
A.~Castellano, A.~Herr{\'a}ez and L.E.~Ib{\'a}{\~n}ez, \emph{{The emergence
  proposal in quantum gravity and the species scale}},
  \href{https://doi.org/10.1007/JHEP06(2023)047}{\emph{JHEP} {\bfseries 06}
  (2023) 047} [\href{https://arxiv.org/abs/2212.03908}{{\ttfamily
  2212.03908}}].

\bibitem{Castellano:2023stg}
A.~Castellano, I.~Ruiz and I.~Valenzuela, \emph{{Universal Pattern in Quantum
  Gravity at Infinite Distance}},
  \href{https://doi.org/10.1103/PhysRevLett.132.181601}{\emph{Phys. Rev. Lett.}
  {\bfseries 132} (2024) 181601}
  [\href{https://arxiv.org/abs/2311.01501}{{\ttfamily 2311.01501}}].

\bibitem{Castellano:2023jjt}
A.~Castellano, I.~Ruiz and I.~Valenzuela, \emph{{Stringy evidence for a
  universal pattern at infinite distance}},
  \href{https://doi.org/10.1007/JHEP06(2024)037}{\emph{JHEP} {\bfseries 06}
  (2024) 037} [\href{https://arxiv.org/abs/2311.01536}{{\ttfamily
  2311.01536}}].

\bibitem{Arkani-Hamed:2006emk}
N.~Arkani-Hamed, L.~Motl, A.~Nicolis and C.~Vafa, \emph{{The String landscape,
  black holes and gravity as the weakest force}},
  \href{https://doi.org/10.1088/1126-6708/2007/06/060}{\emph{JHEP} {\bfseries
  06} (2007) 060} [\href{https://arxiv.org/abs/hep-th/0601001}{{\ttfamily
  hep-th/0601001}}].

\bibitem{Heidenreich:2015nta}
B.~Heidenreich, M.~Reece and T.~Rudelius, \emph{{Sharpening the Weak Gravity
  Conjecture with Dimensional Reduction}},
  \href{https://doi.org/10.1007/JHEP02(2016)140}{\emph{JHEP} {\bfseries 02}
  (2016) 140} [\href{https://arxiv.org/abs/1509.06374}{{\ttfamily
  1509.06374}}].

\bibitem{Cordova:2019jnf}
C.~C{\'o}rdova, D.S.~Freed, H.T.~Lam and N.~Seiberg, \emph{{Anomalies in the
  Space of Coupling Constants and Their Dynamical Applications I}},
  \href{https://doi.org/10.21468/SciPostPhys.8.1.001}{\emph{SciPost Phys.}
  {\bfseries 8} (2020) 001} [\href{https://arxiv.org/abs/1905.09315}{{\ttfamily
  1905.09315}}].

\bibitem{Cordova:2019uob}
C.~C{\'o}rdova, D.S.~Freed, H.T.~Lam and N.~Seiberg, \emph{{Anomalies in the
  Space of Coupling Constants and Their Dynamical Applications II}},
  \href{https://doi.org/10.21468/SciPostPhys.8.1.002}{\emph{SciPost Phys.}
  {\bfseries 8} (2020) 002} [\href{https://arxiv.org/abs/1905.13361}{{\ttfamily
  1905.13361}}].

\bibitem{McNamara:2020uza}
J.~McNamara and C.~Vafa, \emph{{Baby Universes, Holography, and the
  Swampland}},  \href{https://arxiv.org/abs/2004.06738}{{\ttfamily
  2004.06738}}.

\bibitem{Heidenreich:2020pkc}
B.~Heidenreich, J.~McNamara, M.~Montero, M.~Reece, T.~Rudelius and
  I.~Valenzuela, \emph{{Chern-Weil global symmetries and how quantum gravity
  avoids them}}, \href{https://doi.org/10.1007/JHEP11(2021)053}{\emph{JHEP}
  {\bfseries 11} (2021) 053}
  [\href{https://arxiv.org/abs/2012.00009}{{\ttfamily 2012.00009}}].

\bibitem{Witten:1992xu}
E.~Witten, \emph{{Two-dimensional gauge theories revisited}},
  \href{https://doi.org/10.1016/0393-0440(92)90034-X}{\emph{J. Geom. Phys.}
  {\bfseries 9} (1992) 303}
  [\href{https://arxiv.org/abs/hep-th/9204083}{{\ttfamily hep-th/9204083}}].

\bibitem{Ooguri:2006in}
H.~Ooguri and C.~Vafa, \emph{{On the Geometry of the String Landscape and the
  Swampland}},
  \href{https://doi.org/10.1016/j.nuclphysb.2006.10.033}{\emph{Nucl. Phys. B}
  {\bfseries 766} (2007) 21}
  [\href{https://arxiv.org/abs/hep-th/0605264}{{\ttfamily hep-th/0605264}}].

\bibitem{Ooguri:2018wrx}
H.~Ooguri, E.~Palti, G.~Shiu and C.~Vafa, \emph{{Distance and de Sitter
  Conjectures on the Swampland}},
  \href{https://doi.org/10.1016/j.physletb.2018.11.018}{\emph{Phys. Lett. B}
  {\bfseries 788} (2019) 180}
  [\href{https://arxiv.org/abs/1810.05506}{{\ttfamily 1810.05506}}].

\bibitem{Grimm:2018ohb}
T.W.~Grimm, E.~Palti and I.~Valenzuela, \emph{{Infinite Distances in Field
  Space and Massless Towers of States}},
  \href{https://doi.org/10.1007/JHEP08(2018)143}{\emph{JHEP} {\bfseries 08}
  (2018) 143} [\href{https://arxiv.org/abs/1802.08264}{{\ttfamily
  1802.08264}}].

\bibitem{Font:2019cxq}
A.~Font, A.~Herr{\'a}ez and L.E.~Ib{\'a}{\~n}ez, \emph{{The Swampland Distance
  Conjecture and Towers of Tensionless Branes}},
  \href{https://doi.org/10.1007/JHEP08(2019)044}{\emph{JHEP} {\bfseries 08}
  (2019) 044} [\href{https://arxiv.org/abs/1904.05379}{{\ttfamily
  1904.05379}}].

\bibitem{Calderon-Infante:2023ler}
J.~Calder{\'o}n-Infante, A.~Castellano, A.~Herr{\'a}ez and L.E.~Ib{\'a}{\~n}ez,
  \emph{{Entropy bounds and the species scale distance conjecture}},
  \href{https://doi.org/10.1007/JHEP01(2024)039}{\emph{JHEP} {\bfseries 01}
  (2024) 039} [\href{https://arxiv.org/abs/2306.16450}{{\ttfamily
  2306.16450}}].

\bibitem{Hogan-Murphy:lebesgue-integration}
D.~Hogan-Murphy, ``Lebesgue integration.''
  \url{https://math.uchicago.edu/~may/REU2024/REUPapers/Hogan-Murphy.pdf},
  2025.
\newblock Accessed: 2026-01-20.

\bibitem{Jimu:2024xqm}
D.~Jimu and T.~Prokopec, \emph{{Uniqueness of gravitational constant at low
  energies from the connection between spin-2 and spin-0 sectors}},
  \href{https://doi.org/10.1007/JHEP04(2025)134}{\emph{JHEP} {\bfseries 04}
  (2025) 134} [\href{https://arxiv.org/abs/2410.01449}{{\ttfamily
  2410.01449}}].

\end{thebibliography}\endgroup

\end{document}